\documentclass[11pt,a4paper]{article}

\usepackage{graphicx}
\usepackage{float}
\usepackage{afterpage}
\usepackage{epsfig,cite}
\usepackage{amssymb}
\usepackage{amsmath}
\usepackage{dsfont}
\usepackage{multirow}
\usepackage{xspace}
\usepackage{inconsolata}
\usepackage{tabularx}
\usepackage{booktabs}
\usepackage{relsize}
\usepackage{jheppub} 
\usepackage[T1]{fontenc}
\usepackage{titlesec}
\usepackage{listings}
\usepackage{scalefnt}

\usepackage{color}
\usepackage{fancyvrb}

\DefineVerbatimEnvironment{Highlighting}{Verbatim}{commandchars=\\\{\}}
\usepackage{framed}
\definecolor{shadecolor}{RGB}{248,248,248}
\newenvironment{Shaded}{\begin{snugshade}}{\end{snugshade}}

\newcommand{\ExtensionTok}[1]{#1}

\newcommand{\NormalTok}[1]{#1}

\usepackage{url}

\newcommand{\beq}{\begin{equation}}
\newcommand{\eeq}{\end{equation}}
\newcommand{\bea}{\begin{align}}
\newcommand{\eea}{\end{align}}

\newcommand{\as}{\alpha_s}

\newcommand{\pt}{p_T}

\newcommand{\ptWW}{p_{T}^{WW}}
\newcommand{\MWW}{M_{WW}}
\newcommand{\ptj}{p_T^J}
\newcommand{\eps}{\delta}
\newcommand{\dZ}{\rd\mathcal{Z}[\{{\tildeR}', k_i\}]}

\newcommand{\ptjv}{{p_T^{J,{\rm veto}}}}
\newcommand{\kto}{k_{t1}}

\newcommand{\tildeL}{{L}}
\newcommand{\tildecalL}{{\cal L}}
\newcommand{\tildeR}{R}
\newcommand{\tildeC}{C}
\newcommand{\tildeH}{H}
\newcommand{\ptlep}{\ensuremath{p_{T,{\ell}}}}

\newcommand{\etalep}{\ensuremath{|\eta_{\ell}|}}

\newcommand{\mll}{\ensuremath{m_{\ell^-\ell^+}}}
\newcommand{\ptmiss}{\ensuremath{p_{T}^{\text{miss}}}}
\makeatletter
\g@addto@macro\bfseries{\boldmath}
\makeatother

\allowdisplaybreaks

\DeclareRobustCommand{\ensuremathrm}[1]{\ensuremath{\mathrm{#1}}\xspace}

\DeclareRobustCommand{\phs}{\ensuremath{\phi^{*}_\eta}\xspace}
\DeclareRobustCommand{\rd}{\ensuremathrm{d}} 
\DeclareRobustCommand{\re}{\ensuremathrm{e}} 

\usepackage[scr=rsfso]{mathalfa}
\usepackage{bm}
\usepackage{inconsolata}

\usepackage{xcolor}
\definecolor{darkgreen}{rgb}{0,0.6,0}
\definecolor{darkpurple}{rgb}{0,0.5,0.5}
\definecolor{darkblue}{rgb}{0,0,0.7}
\definecolor{darkred}{rgb}{0.5,0,0.0}
\definecolor{darkorange}{rgb}{0.8,0.4,0.0}
\definecolor{green}{rgb}{0.0,0.8,0.4}

\titleformat{\paragraph}
{\normalfont\normalsize\bfseries}{\theparagraph}{1em}{}
\titlespacing*{\paragraph}
{0pt}{3.25ex plus 1ex minus .2ex}{1.5ex plus .2ex}

\preprint{
  \begin{flushright}
    MPP-2020-53\\
    LAPTH-012/20
  \end{flushright}
}

\begin{document}

\title{Accurate single- and double-differential resummation of colour-singlet processes
with M{\scalefont{0.8}ATRIX}+R{\scalefont{0.8}AD}ISH: $W^+W^-$ production at the LHC }

\author[1]{Stefan~Kallweit,}
\author[2]{Emanuele~Re,}
\author[1,3]{Luca~Rottoli}
\author[4]{and Marius~Wiesemann}

\affiliation[1]{Dipartimento di Fisica G. Occhialini, Universit\`a degli Studi di Milano-Bicocca and INFN, Piazza della Scienza 3, 20126 Milano, Italy}
\affiliation[2]{LAPTh, Universit\'e Grenoble Alpes, Universit\'e Savoie Mont Blanc, CNRS, 74940 Annecy, France}
\affiliation[3]{Ernest Orlando Lawrence Berkeley National Laboratory, University of California, Berkeley, CA 94720, USA}
\affiliation[4]{Max-Planck Institut f\"ur Physik, F\"ohringer Ring 6, 80805 M\"unchen, Germany }

\emailAdd{stefan.kallweit@unimib.it}
\emailAdd{emanuele.re@lapth.cnrs.fr}
\emailAdd{luca.rottoli@unimib.it}
\emailAdd{marius.wiesemann@cern.ch} 

\abstract{
We present the combination of fully differential cross sections for colour-singlet production processes at next-to-next-to-leading order (NNLO) QCD obtained with \textsc{Matrix} and all-order resummation through \textsc{RadISH}. This interface allows us to achieve unprecedented accuracy for various transverse observables in \mbox{$2 \rightarrow 2$} production processes.
As an important application we consider $W^+ W^-$ production at the LHC, more precisely the full leptonic process $pp\to \ell^+\ell^{\prime\, -}\nu_{\ell}{\bar\nu}_{\ell^\prime}+X$ with $\ell'\neq\ell$, and we present resummed predictions for differential distributions in presence of fiducial selection cuts.
In particular, we resum the transverse-momentum spectrum of the $W^+ W^-$ pair at next-to-next-to-next-to-leading logarithmic (N$^3$LL) accuracy and match it to the integrated NNLO cross section. The transverse-momentum spectrum of the leading jet in $W^+ W^-$ production is calculated at NNLO+NNLL accuracy. Finally, the joint resummation
for the transverse-momentum spectrum of the $W^+ W^-$ pair in the presence of a jet veto is performed at NNLO+NNLL.
Our phenomenological study highlights the importance of 
higher-order perturbative and logarithmic corrections
for precision phenomenology at the LHC.
}

\maketitle

\section{Introduction}

Precision phenomenology has become of major importance in the rich physics programme at the Large Hadron Collider (LHC). 
Lacking any direct observations of physics beyond the SM (BSM), precision measurements 
provide a valuable alternative in the 
discovery of new-physics phenomena through small deviations from the Standard Model (SM) picture.
The vast amount of data collected at the LHC continuously decreases experimental uncertainties, 
thereby demanding accurate predictions for various observables in many relevant physics processes.

Differential distributions in colour-singlet production processes and associated QCD radiation play a special role in this context.
Being measured by reconstructing the colour singlet from its leptonic decay products, these processes usually provide very clean experimental signatures. Therefore, such 
processes are often characterised by particularly small experimental uncertainties, which in the case of 
Drell--Yan production can reach sub-percent precision and are at the few-percent level even for vector-boson pair production processes.
Due to their precise measurement, leptonic processes induced by vector-boson decays provide prime signatures to
extract SM parameters, to constrain parton densities, 
or to calibrate event-generation tools used in experimental analyses.
Not least, vector-boson processes have a high sensitivity
to new-physics phenomena, and they can be exploited to
set stringent limits on BSM effects, which induce shape distortions in kinematic distributions. 

Precision measurements require predictions that match the experimental uncertainties to fully exploit the vast potential of LHC data.
The theoretical description of fiducial cross sections and kinematic distributions has been greatly improved by the calculation of NNLO QCD corrections, which are nowadays necessary to reliably describe the experimental data away from the soft and collinear regions. 
Fully differential predictions at NNLO QCD in the Born kinematics are available by now for essentially all
\mbox{$2 \rightarrow 2$} colour-singlet production processes, including $HZ$ and $HW^\pm$ production~\cite{Ferrera:2013yga,Ferrera:2014lca,Campbell:2016jau,Ferrera:2017zex,Caola:2017xuq,Gauld:2019yng},  $HH$ production~\cite{deFlorian:2013jea,deFlorian:2016uhr,Grazzini:2018bsd}, $\gamma\gamma$ production~\cite{Catani:2011qz,Campbell:2016yrh,Catani:2018krb}, $Z\gamma$ and $W^\pm\gamma$ production~\cite{Grazzini:2013bna,Grazzini:2015nwa,Campbell:2017aul}, $W^+ W^-$ production~\cite{Gehrmann:2014fva,Grazzini:2016ctr}, $ZZ$ production~\cite{Cascioli:2014yka,Grazzini:2015hta,Heinrich:2017bvg,Kallweit:2018nyv} and $W^{\pm}Z$ production~\cite{Grazzini:2016swo,Grazzini:2017ckn}. 

It is well known that in kinematical regimes dominated by soft and collinear QCD radiation the perturbative expansion in the strong coupling constant $\as$ does not provide a physical description of differential observables. 
An important class of distributions is that of \textit{transverse} observables, which do not depend on the rapidity of the radiation.
For colour-singlet processes prime examples of such observables are the transverse momentum  
of the colourless final state $\pt$ or of the leading accompanying jet $\ptj$.
In phase space regions dominated by soft and collinear radiation, the perturbative expansion of the cross section 
is marred by large logarithms 
$L = \ln (v)$,
where the symbol $v$ denotes a general transverse observable, such as $\pt$ or $\ptj$.
A resummation of these logarithmically-enhanced terms to all orders in $\as$ is required to obtain physical results when $v$ becomes small.
The logarithmic accuracy is customarily defined in terms of the logarithm of the cumulative cross section $\ln \sigma (v)$.
One refers to the dominant terms $\as^n L^{n+1}$ as leading logarithmic (LL), to terms $\as^n L^{n}$ as next-to-leading logarithmic (NLL), to $\as^n L^{n-1}$ as next-to-next-to-leading logarithmic (NNLL), and so on.

A variety of formalisms to perform the resummation of large logarithmic contributions for transverse observables in colour-singlet processes 
has been developed over the last four decades~\cite{Parisi:1979se,Collins:1984kg,Balazs:1995nz,Ellis:1997sc,Balazs:1997xd,Idilbi:2005er,Bozzi:2005wk,Catani:2010pd,Becher:2010tm,Becher:2011xn,GarciaEchevarria:2011rb,Banfi:2012yh,Becher:2012qa,Banfi:2012jm,Banfi:2012du,Chiu:2012ir,Becher:2013xia,Stewart:2013faa,Neill:2015roa,Monni:2016ktx,Ebert:2016gcn,Bizon:2017rah}.
The most accurate description of the $\pt$ spectrum has been achieved so far for Higgs boson production and for the neutral and charged 
Drell--Yan production in refs.~\cite{Bizon:2017rah,Chen:2018pzu,Bizon:2018foh,Bizon:2019zgf}, with N$^3$LL resummation matched to NNLO QCD predictions for $F$+jet production of refs.~\cite{Caola:2015wna,Gehrmann-DeRidder:2016jns,Chen:2016zka,Gehrmann-DeRidder:2017mvr}, where $F$ denotes the respective colour singlet \footnote{Note that ref.~\cite{Becher:2019bnm} also performed N$^3$LL resummation for neutral Drell-Yan production, but only matched it to NLO QCD predictions for $Z$+jet production.}.
Similarly, the resummation for $\ptj$ has been calculated at NNLL accuracy~\cite{Becher:2012qa,Banfi:2012jm,Becher:2013xia,Stewart:2013faa}, and it has been used to produce predictions for jet-vetoed cross sections up to N$^3$LO+NNLL accuracy in Higgs boson production~\cite{Banfi:2015pju}.
Very recently, a formalism to simultaneously resum both classes of logarithms at NNLL accuracy has been presented in ref.~\cite{Monni:2019yyr}, where it was 
applied to obtain the Higgs $\pt$ spectrum with a jet veto at NNLL matched to the NLO prediction for $H$+jet production.

Although these resummation formalisms can be generalised to the production of an arbitrary colourless final state, there is 
only a limited number of calculations that have included high-accuracy resummation in \mbox{$2 \rightarrow 2$} processes due to their higher complexity.
Resummed results for the transverse momentum of 
$HH$ \cite{Ferrera:2016prr}, $\gamma \gamma$~\cite{Cieri:2015rqa}, $W^+W^-$~\cite{Meade:2014fca,Grazzini:2015wpa}, $ZZ$~\cite{Grazzini:2015wpa}, and $W^\pm Z$~\cite{Becher:2019bnm} have been computed up to NNLL accuracy, while jet-veto logarithms have been resummed at NNLL accuracy in $HZ$ and $HW^\pm$ production~\cite{Shao:2013uba,Li:2014ria}, $W^+ W^-$~\cite{Becher:2014aya,Dawson:2016ysj,Arpino:2019fmo} production, and $ZZ$ production~\cite{Arpino:2019fmo}.
Note, however, that refs.~\cite{Cieri:2015rqa,Grazzini:2015wpa,Dawson:2016ysj} were the only ones of those which matched to the NNLO cross section, whereas the 
other studies used less accurate fixed-order predictions.\footnote{Here and in what follows the fixed-order accuracy refers to the one of the integrated cross section, not of the spectrum. Thus, matching to the (N)NLO cross section implies that accuracy only after integration over the resummed observable, and (N)LO accuracy in the spectrum (or accordingly for the $F$+jet process).}
In most of the available resummation codes, such as \textsc{arTeMiDe}~\cite{Bertone:2019nxa}, \textsc{CuTe}~\cite{Becher:2012yn}, \textsc{DYRES/DYTURBO}~\cite{Catani:2015vma,Camarda:2019zyx}, \textsc{HqT}~\cite{Bozzi:2005wk}, \textsc{HRES}~\cite{deFlorian:2012mx}, \textsc{NangaParbat}~\cite{Bacchetta:2019sam}, \textsc{RadISH}~\cite{Bizon:2017rah}, \textsc{ResBOS}~\cite{Balazs:1997xd} and \textsc{Resolve}~\cite{Coradeschi:2017zzw},  only a few colour-singlet production processes are available, including in most cases only
Higgs or Drell--Yan production.
Notable exceptions are the \textsc{Matrix} code~\cite{Grazzini:2015wpa,Grazzini:2017mhc}
and the framework of ref.~\cite{Becher:2019bnm}, which implement
transverse-momentum resummation for several colour-singlet processes at NNLL accuracy, matching to the NNLO (NLO) cross section in the case of \textsc{Matrix} (the framework of ref.~\cite{Becher:2019bnm}).
Also for jet-veto resummation more general frameworks have been 
developed~\cite{Becher:2014aya,Arpino:2019fmo}, which evaluate jet-vetoed cross sections at NLO+NNLL accuracy for various production processes of electroweak~(EW) bosons. 
Nonetheless, currently there is no unique framework which is sufficiently flexible to resum various observables (simultaneously) at state-of-the-art accuracy matched 
to NNLO QCD predictions for arbitrary colour-singlet final states.

In this paper, we present the \textsc{Matrix+RadISH} framework for high-accuracy resummed calculations at the multi-differential level. 
By developing a general interface between the \textsc{Matrix} and the \textsc{RadISH} codes, a substantial advancement 
over previous resummed predictions is achieved for a large number of non-trivial colour-singlet processes at the LHC. 
All the \mbox{$2 \rightarrow 1$} and \mbox{$2 \rightarrow 2$} processes available in the public release of \textsc{Matrix} are included, and 
essentially any colour-singlet process for which two-loop amplitudes become available can be added. 
Through its powerful and versatile parton-level Monte Carlo generator, \textsc{Matrix+RadISH} provides an accurate description of several transverse observables in colour-singlet processes.
In particular, it facilitates transverse-momentum resummation of the colour-singlet final state at N$^3$LL accuracy,
$\phs$~\cite{Banfi:2010cf} resummation for the Drell--Yan process at N$^3$LL accuracy, 
transverse-momentum resummation of the leading jet (and equivalently jet-veto resummation) at NNLL accuracy,
and double-differential resummation in the transverse momentum of the colour singlet and of the leading jet at NNLL accuracy.
The latter allows for the consistent calculation of the $\pt$ ($\ptj$) spectrum with a veto cut on $\ptj$ ($\pt$). The matching is performed to the integrated cross section at NNLO QCD accuracy.

The resummation is formulated in the \textsc{RadISH} formalism of refs.~\cite{Monni:2016ktx,Bizon:2017rah}, which allows us to resum transverse observables at high accuracy.
The \textsc{RadISH} code also contains the implementation of the double-differential resummation on the basis of ref.~\cite{Monni:2019yyr}.
The fixed-order component, the phase space and the relevant perturbative ingredients are evaluated through the computational 
framework \textsc{Matrix}~\cite{Grazzini:2015wpa,Grazzini:2017mhc}.

As a first phenomenological application we study the production of $W^+W^-$ pairs in hadronic collisions.  More precisely, we consider the 
full leptonic process with two charged leptons of different flavour and the two corresponding neutrinos in the final state. By evaluating 
all resonant and non-resonant contributions we include off-shell effects and spin correlations.
In this paper, we advance the current state of the art of predictions in terms of accuracy for various observables 
at the multi-differential level in $W^+ W^-$ production. In particular, we compute the fiducial cross section as a function of 
the jet-veto cut at NNLO+NNLL accuracy, and compare our predictions with recent ATLAS data~\cite{Aaboud:2019nkz}.
Furthermore, we calculate fiducial predictions for the $\ptj$ spectrum at NNLO+NNLL accuracy and 
for the transverse momentum of the $W^+ W^-$ pair at NNLO+N$^3$LL accuracy.
Finally, we present double-differentially resummed predictions for the transverse-momentum spectrum of the $W^+ W^-$ pair 
at NNLO+NNLL accuracy in presence of a veto on $\ptj$.

The manuscript is organised as follows: In section~\ref{sec:framework} we give a general introduction to \textsc{Matrix+RadISH}.
In particular, we review the computation of NNLO corrections for colour-singlet production within \textsc{Matrix} (section~\ref{sec:matrix}), provide the relevant formul\ae\ for the resummation 
of transverse observables in the \textsc{RadISH} formalism  (section~\ref{sec:radish}), discuss their matching (section~\ref{sec:matching}), and give details on the practical implementation (section~\ref{sec:implementation}).
In section~\ref{sec:WWprod} we discuss the case of $W^+ W^-$ production at the LHC, and 
we present results for fiducial predictions of transverse observables 
both at the single-differential and at the double-differential level.
The main results are summarised in section~\ref{sec:conclusion},
and we provide a practical description of how to use the \textsc{Matrix+RadISH} interface in appendix~\ref{app:example}.


\section{Description of \textsc{Matrix+RadISH}}\label{sec:framework}

In this section, we discuss the calculation of all-order resummation matched to fixed-order predictions with \textsc{Matrix+RadISH}. Our implementation is completely general,
and it can be applied 
to essentially any colour-singlet process. The combination of \textsc{Matrix} and \textsc{RadISH} facilitates the calculation of consistently resummed and matched predictions 
for several observables. The results are accurate up to NNLO in QCD 
perturbation theory for the
fully differential cross section of the produced colour-singlet final state.
In particular, the framework allows us to evaluate the following resummed predictions at unprecedented precision for $2\rightarrow 2$
colour-singlet production processes:
\begin{itemize}
\item single-differential resummation for the transverse-momentum spectrum 
of a colour-singlet ($p_T$) up to NNLO+N$^3$LL accuracy,
\item single-differential resummation for the $\phs$ distribution in the Drell--Yan process up to NNLO+N$^3$LL accuracy,
\item single-differential resummation
for the transverse-momentum distribution of the leading jet ($\ptj$) (and equivalently for the jet-vetoed cross section) up to NNLO+NNLL accuracy,
\item double-differential resummation of $p_T$ and $\ptj$ logarithms up to NNLO+NNLL accuracy, allowing us to evaluate the $p_T$ ($\ptj$) spectrum with a veto on $\ptj$ ($p_T$) at NNLO+NNLL accuracy.
\end{itemize}
The calculations are fully differential in all Born level observables, and arbitrary fiducial cuts can be applied to the final state particles.\footnote{Note that isolation cuts between QCD radiation and leptons or photons may introduce additional logarithmic corrections of non-global nature~\cite{Balsiger:2018ezi}.}

\subsection{Higher-order corrections with \textsc{Matrix}}\label{sec:matrix}

\textsc{Matrix} is a general framework for fixed-order calculations in QCD and EW perturbation theory, covering a 
large number of primary LHC scattering processes. The public release of \textsc{Matrix}~\cite{Grazzini:2017mhc,Matrixurl} evaluates NNLO QCD 
predictions for $2\to 1$ and $2\to 2$ colour-singlet processes~\cite{Grazzini:2013bna,Grazzini:2015nwa,Cascioli:2014yka,Grazzini:2015hta,Gehrmann:2014fva,Grazzini:2016ctr,Grazzini:2016swo,Grazzini:2017ckn,Kallweit:2018nyv}\footnote{\textsc{Matrix} has also been applied to calculate NNLO QCD cross sections for Higgs boson pair production~\cite{deFlorian:2016uhr,Grazzini:2018bsd} and Higgsstrahlung~\cite{Alioli:2019qzz}, and it has been recently extended to top-quark pair production~\cite{Catani:2019iny,Catani:2019hip}.}, including
all possible leptonic decay channels of the massive vector bosons, while consistently accounting for resonant and 
non-resonant diagrams, off-shell effects and spin correlations. More recently, \textsc{Matrix} predictions have been further advanced by including 
important effects beyond NNLO QCD: The dominant next-to-NNLO (N$^3$LO) QCD corrections have been implemented by calculating 
the loop-induced gluon fusion contribution at NLO QCD for $ZZ$~\cite{Grazzini:2018owa} and $W^+W^-$~\cite{Grazzini:2020stb} production, and 
the combination of NNLO QCD and NLO EW corrections has been achieved for all the leptonic final states of massive diboson processes~\cite{Kallweit:2019zez}.\footnote{\textsc{Matrix} was also used in the NNLO+NNLL computation of ref.~\cite{Grazzini:2015wpa} and in the NNLOPS computation of ref.~\cite{Re:2018vac}.}
A new release of \textsc{Matrix} with these corrections is currently in preparation.

In this paper, we make use of the general implementation of fully differential NNLO cross sections
in QCD perturbation theory for colour-singlet processes within \textsc{Matrix}.
The computation of NNLO QCD corrections requires 
the evaluation of tree-level contributions with zero, one and two additional partons,
of one-loop contributions with zero and one parton and of purely virtual two-loop contributions.
Their combination in a fully differential (exclusive) calculation at NNLO QCD is 
highly non-trivial since infrared~(IR) divergences affect real and virtual contributions in different ways, 
preventing a straightforward combination of these components.
To overcome these issues, \textsc{Matrix} features a fully general implementation of the $q_T$-subtraction method~\cite{Catani:2007vq} at NNLO QCD, which is briefly described below.
In this context, an automated extrapolation procedure has been implemented to calculate integrated 
cross sections in the limit in which the $q_T$-subtraction cutoff parameter goes to zero~\cite{Grazzini:2017mhc}.
The core of the \textsc{Matrix} framework \cite{Grazzini:2017mhc} is the Monte Carlo program \textsc{Munich}\footnote{\textsc{Munich} is the 
abbreviation of ``MUlti-chaNnel Integrator at Swiss~(CH) precision'' --- an automated parton-level
NLO generator by S.~Kallweit.}, which is capable of
computing both NLO QCD and NLO EW~\cite{Kallweit:2014xda,Kallweit:2015dum} corrections
to arbitrary SM processes.
All tree-level and one-loop amplitudes are supplied by \textsc{OpenLoops}~\cite{Cascioli:2011va,Buccioni:2017yxi,Buccioni:2019sur} through an automated interface.
For validation and stability tests of the employed amplitudes a similar interface to the \textsc{Recola} amplitude generator~\cite{Actis:2016mpe,Denner:2017wsf} has been implemented.
At two-loop level, for massive diboson production the public C++ library \textsc{VVamp}~\cite{hepforge:VVamp} is used 
that implements the $q\bar{q}\to VV'$ and $gg\to VV'$ helicity amplitudes of refs.~\cite{Gehrmann:2015ora,vonManteuffel:2015msa}.\footnote{Results for these
  two-loop amplitudes have been evaluated independently in refs.~\cite{Caola:2014iua,vonManteuffel:2015msa}.} For $V\gamma$~\cite{Gehrmann:2011ab} and $\gamma\gamma$~\cite{Anastasiou:2002zn} production we rely on private
implementations of the respective amplitudes.

The $q_T$-subtraction formalism \cite{Catani:2007vq} exploits the fact that 
the behaviour of the cross section at small transverse momentum of a colour-singlet final-state system
has a universal (process-independent) structure that is explicitly known up to NNLO QCD
through the formalism of transverse-momentum resummation \cite{Collins:1984kg,Bozzi:2005wk}.
This knowledge is sufficient to construct a non-local, but process-independent IR subtraction counterterm for 
this entire class of processes. In the $q_T$-subtraction method, the NNLO cross section for a general process 
$pp\to F+X$, where $F$ is a colourless system, is written as
\begin{equation}
\label{eq:qtsub}
{\rm d}{\sigma}^{F}_{\mathrm{NNLO}}=\left[ {\rm d}{\sigma}^{F\mathrm{+ jet}}_{\mathrm{NLO}}-
{\rm d}{\sigma}^{\mathrm{CT}}_{\mathrm{NNLO}}\right] + {\cal H}^{F}_{\mathrm{NNLO}}\otimes {\rm d}{\sigma}^{F}_{\mathrm{LO}},
\end{equation}
where $d{\sigma}^{F\mathrm{ + jet}}_{\mathrm{NLO}}$ is the cross section for the production of the system $F$ and a jet at NLO accuracy, which can be 
evaluated by using one of the available NLO subtraction methods \cite{Frixione:1995ms,Frixione:1997np,Catani:1996jh,Catani:1996vz} to
cancel the corresponding IR divergencies. In fact, unless the transverse-momentum of the colour singlet approaches zero, $d{\sigma}^{F\mathrm{ + jet}}_{\mathrm{NLO}}$ is finite.
The process-independent counterterm  ${\rm d}{\sigma}^{\mathrm{CT}}_{\mathrm{NNLO}}$ cancels the remaining divergence 
in the limit of vanishing transverse momentum, and it is constructed by expanding the transverse-momentum resummation formula \cite{Collins:1984kg,Bozzi:2005wk}
up to NNLO. The computation is completed by the last term on the right-hand side of eq.~\eqref{eq:qtsub} that
depends on the hard-collinear function ${\cal H}^{F}_{\mathrm{NNLO}}$ up to NNLO~\cite{deFlorian:2001zd,Catani:2013tia,Catani:2011kr,Catani:2012qa}.

The practical implementation of the $q_T$-subtraction formalism in \textsc{Matrix} deserves some additional discussion.
The contribution in the square bracket in eq.~\eqref{eq:qtsub} is formally finite,
but each individual term ${\rm d}{\sigma}^{F\mathrm{ + jet}}_{\mathrm{NLO}}$ and ${\rm d}{\sigma}^{\mathrm{CT}}_{\mathrm{NNLO}}$
is separately divergent. Since the subtraction is not local,
a technical cut-off $r_{\mathrm{cut}}$ on the dimensionless quantity $r=p_T/M$, where $p_T$ is the transverse momentum and $M$ is the invariant mass of the colourless system, is introduced, rendering both terms separately finite. 
Below this cut-off ${\rm d}{\sigma}^{F\mathrm{ + jet}}_{\mathrm{NLO}}$ 
and ${\rm d}{\sigma}^{\mathrm{CT}}_{\mathrm{NNLO}}$ are assumed to be identical, which is correct up to power-suppressed terms.
These power-suppressed terms vanish only in the limit $r_{\mathrm{cut}}\to0$, but their impact is controlled by monitoring the dependence of the cross section on $r_{\mathrm{cut}}$.
To this end, \textsc{Matrix} simultaneously computes the cross section at several $r_{\mathrm{cut}}$ values without the need of repeated CPU-intensive runs,
which is used to extrapolate the cross section in the $r_{\mathrm{cut}}\to0$ limit by fitting the results at finite $r_{\mathrm{cut}}$ values.
The extrapolated result and an estimate of the respective uncertainty are provided at the end of every \textsc{Matrix} run.
We note that the $q_T$-subtraction method works very similar to a phase space slicing method in this way, with $r_{\mathrm{cut}}$ acting as a slicing parameter.

To perform the resummation of large logarithmic contributions, we have implemented a general interface to combine the \textsc{Matrix} framework with the \textsc{RadISH} code, which is introduced in the next section.
In this context, \textsc{Matrix} provides all the fixed-order parts of the calculation as well as the Born level phase space points and the hard coefficients 
needed for the calculation of the resummed component.
The latter are passed to \textsc{RadISH} which evaluates the 
resummation for the observable under consideration.
In this paper we present a detailed study of the phenomenological implications 
for $W^+W^-$ production only, but our implementation is completely general, and can directly be used for any of the other colour-singlet processes available in \textsc{Matrix}.

\subsection{Resummation of large logarithmic contributions with \textsc{RadISH}}\label{sec:radish}

The \textsc{RadISH} approach for the resummation of transverse observables in colour-singlet processes has been presented in refs.~\cite{Monni:2016ktx,Bizon:2017rah} 
and will be summarised in the following. By exploiting the factorisation properties of squared QCD amplitudes and the recursive infrared collinear safety (rIRC)~\cite{Banfi:2004yd} of the considered observables the resummation is formulated
directly in momentum space, thereby obtaining a more differential description of the QCD radiation than that in customary conjugate-space formulations.
The resummation is numerically evaluated via efficient Monte Carlo methods, yielding a powerful formalism similar in spirit to a semi-inclusive parton shower, but with 
the consistent inclusion of higher-order logarithmic contributions and full control over the formal accuracy.
Thanks to its versatility, the approach can be exploited to formulate the resummation for the entire class of transverse observables, i.e.\ those which do not depend on the rapidity of the radiation, in a unique framework.
This enabled the recent extension to 
double-differential resummation of the transverse-momentum spectrum of the colour singlet with a jet veto in ref.~\cite{Monni:2019yyr}, as we will discuss below. 

The~\textsc{RadISH} formul\ae\ for the resummation of transverse observables are conveniently expressed in terms of the cumulative cross section
\begin{equation}
\label{eq:cumulative}
\sigma(v) \equiv \int_0^{v} \rd v'\; \frac{\rd \sigma(v')}{\rd v'}\,,
\end{equation}
where the transverse observable $v'=V(\Phi_B,k_1,\dots,k_n)$ is a function of the Born phase space~$\Phi_B$ of the produced colour singlet
and of the momenta $k_1,...,k_n$ of $n$ real emissions. 
The all-order structure of the cumulative distribution $\sigma(v)$, differential in the Born phase space, can be expressed as
\begin{equation}
  \label{eq:sigma-2}
  \frac{\rd\sigma(v)}{\rd \Phi_B} =  {\cal V}(\Phi_B) \sum_{n=0}^{\infty}
  \int\prod_{i=1}^n [\rd k_i]
  |{\cal M}(\Phi_B,k_1,\dots ,k_n)|^2\,\Theta\left(v-V(\Phi_B,k_1,\dots,k_n)\right)\,,
\end{equation}
where ${\cal M}$ denotes the matrix element for $n$ real emissions and ${\cal V}(\Phi_B)$ is the resummed form factor that encodes the purely virtual corrections~\cite{Dixon:2008gr}.
The phase spaces of the $i$-th emission $k_i$ and of the Born configuration are denoted by $[\rd k_i]$ and $\rd\Phi_B$, respectively.

For observables which fulfil rIRC safety it is possible to establish a well-defined logarithmic counting in the squared amplitude, thereby providing a systematic way to identify the contributions that enter at a given logarithmic order~\cite{Banfi:2004yd,Banfi:2014sua}.
This is achieved by decomposing the squared amplitude defined in eq.~\eqref{eq:sigma-2} in $n$-particle-correlated blocks containing the correlated portion of the squared $n$-emission soft amplitude and its virtual corrections~\cite{Bizon:2017rah} such that blocks with $n$ particles start contributing one logarithmic order higher than blocks with $(n-1)$ particles. 

The rIRC safety of the observables is further exploited to ensure that the divergences of virtual origin, contained in the $\mathcal V(\Phi_B) $ factor of eq.~\eqref{eq:sigma-2}, cancel those appearing at all perturbative orders in the real matrix elements.
Indeed, the rIRC safety of the observable allows us to introduce a resolution scale $q_0$ on the transverse momentum of the radiation such that
neglecting radiation softer than $q_0$ in the computation of $V(\Phi_B,k_1,\dots,k_n)$ only introduces terms suppressed by powers of $q_0$.
This unresolved radiation can thus be neglected when computing $V(\Phi_B,k_1,\dots,k_n)$, and it can be exponentiated to cancel the divergences of virtual origin at all orders. 
Resolved radiation, i.e. radiation harder than $q_0$, must instead be generated exclusively
since it is constrained by the measurement function $\Theta\left(v-V(\Phi_B,k_1,\dots,k_n)\right)$ in eq.~\eqref{eq:sigma-2}.
The rIRC safety of the observable also ensures that the limit $q_0\to0$ can be taken safely since the dependence of the results upon the resolution scale is power-like.

The discussion above is completely general, and it can be applied to any transverse observable.
We start by discussing a particular class of observables, that of \textit{inclusive} observables, which depend solely on the total transverse momentum of QCD radiation.
For clarity, we will consider the case of the transverse momentum of a general colour singlet. Nevertheless, the same formul\ae\ can be applied to any inclusive transverse observable such as the $\phs$ angle in Drell--Yan production~\cite{Banfi:2010cf}, 
which was resummed at N$^3$LL accuracy in ref.~\cite{Bizon:2018foh}. 
All the ingredients for the N$^3$LL $\pt$ resummation have been computed in refs.~\cite{Catani:2011kr,Catani:2012qa,Gehrmann:2014yya,Echevarria:2016scs,Li:2016ctv,Vladimirov:2016dll,Moch:2018wjh,Lee:2019zop}.
In the \textsc{RadISH} formalism, the resummation is numerically evaluated by setting the resolution scale $q_0$ to a small fraction $\eps > 0$ of the transverse momentum of the block with largest $k_t$, henceforth denoted by $k_{t1}$.  
As a result, the cumulative cross section in momentum space at N$^3$LL accuracy for the production of a colour singlet of mass $M$, fully differential in the Born variables, reads~\cite{Bizon:2017rah}
\begin{align}
\label{eq:master-kt-space}
&\frac{\rd\sigma(\pt)}{\rd\Phi_B} \!  = \!  \! \int\frac{\rd k_{t1}}{k_{t1}}\frac{\rd
  \phi_1}{2\pi}\partial_{\tildeL}\left(-\re^{-\tildeR(k_{t1})} {\tildecalL}_{\rm
  N^3LL}(k_{t1}) \right) \int \dZ\Theta\! \left(\pt-  |\vec{k}_{t,1}+\dots+\vec{k}_{t,n+1}|\right)
                             \notag\\
& + \int\frac{\rd k_{t1}}{k_{t1}}\frac{\rd
  \phi_1}{2\pi} \re^{-\tildeR(k_{t1})} \!  \int \dZ\int_{0}^{1}\frac{\rd \zeta_{s}}{\zeta_{s}}\frac{\rd
  \phi_s}{2\pi}\Bigg\{\bigg({\tildeR}' (k_{t1}) {\tildecalL}_{\rm
  NNLL}(k_{t1}) - \partial_{\tildeL} {\tildecalL}_{\rm
  NNLL}(k_{t1})\bigg) \times\notag\\
&\left(\! {\tildeR}'' (k_{t1}) \!  \ln \! \frac{1}{\zeta_s}\!  + \! \frac{1}{2} {\tildeR}'''
  (k_{t1})\ln^2 \! \frac{1}{\zeta_s} \!  \right)\! -\!  {\tildeR}' (k_{t1})\! \left( \! \partial_{\tildeL} {\tildecalL}_{\rm
  NNLL}(k_{t1})\!  - \!  2\frac{\beta_0}{\pi}\as^2(k_{t1}) \hat{P}^{(0)}\!  \otimes \! {\tildecalL}_{\rm
  NLL}(k_{t1})\!  \ln \! \frac{1}{\zeta_s} \! 
\right)\notag\\
&+\frac{\as^2(k_{t1}) }{\pi^2}\hat{P}^{(0)}\otimes \hat{P}^{(0)}\otimes {\tildecalL}_{\rm
  NLL}(k_{t1})\Bigg\} \bigg\{\Theta\left(\pt-|\vec{k}_{t,1}+\dots+\vec{k}_{t,n+1}+\vec{k}_{t,s}|\right)\notag \\
  & - \Theta\left(\pt-|\vec{k}_{t,1}+\dots+\vec{k}_{t,n+1}|\right)\bigg\}\notag\\
& + \frac{1}{2}\int\frac{\rd k_{t1}}{k_{t1}}\frac{\rd
  \phi_1}{2\pi} \re^{-\tildeR(k_{t1})} \int \dZ\int_{0}^{1}\frac{\rd \zeta_{s1}}{\zeta_{s1}}\frac{\rd
  \phi_{s1}}{2\pi}\int_{0}^{1}\frac{\rd \zeta_{s2}}{\zeta_{s2}}\frac{\rd
  \phi_{s2}}{2\pi} {\tildeR}' (k_{t1})\\
&\times\Bigg\{ {\tildecalL}_{\rm
  NLL}(k_{t1}) \left({\tildeR}'' (k_{t1})\right)^2\ln\frac{1}{\zeta_{s1}} \ln\frac{1}{\zeta_{s2}} - \partial_{\tildeL} {\tildecalL}_{\rm
  NLL}(k_{t1}) {\tildeR}'' (k_{t1})\bigg(\ln\frac{1}{\zeta_{s1}}
  +\ln\frac{1}{\zeta_{s2}} \bigg)\notag\\
&+ \frac{\as^2(k_{t1}) }{\pi^2}\hat{P}^{(0)}\otimes \hat{P}^{(0)}\otimes {\tildecalL}_{\rm
  NLL}(k_{t1})\Bigg\}\notag\\
&\times \bigg\{\Theta\left(\pt-|\vec{k}_{t,1}+\dots+\vec{k}_{t,n+1}+\vec{k}_{t,s1}+\vec{k}_{t,s2}|\right) - \Theta\left(\pt-|\vec{k}_{t,1}+\dots+\vec{k}_{t,n+1}+\vec{k}_{t,s1}|\right) -\notag\\ &\Theta\left(\pt-|\vec{k}_{t,1}+\dots+\vec{k}_{t,n+1}+\vec{k}_{t,s2}|\right) + \Theta\left(\pt-|\vec{k}_{t,1}+\dots+\vec{k}_{t,n+1}|\right)\bigg\}\!  +\!  {\cal O}\left(\as^n \ln^{2n -
                                    6}\frac{1}{v}\right),\notag
\end{align}
where the first line contains the full NLL correction, the first set of curly brackets 
(second to fifth line) 
starts contributing at NNLL accuracy, and the second set of curly brackets 
(from line six)
is a pure N$^3$LL correction.

The luminosity factors $\tildecalL$ in eq.~\eqref{eq:master-kt-space} are evaluated at different orders, and they involve the parton luminosities, the process-dependent squared Born amplitude and hard-virtual corrections $\tildeH^{(n)}$, and the coefficient functions $\tildeC_{ci}^{(n)}$, which have been evaluated to second order
for $gg$- and $q \bar q$-initiated processes in refs.~\cite{Catani:2011kr,Catani:2012qa,Gehrmann:2014yya,Echevarria:2016scs}. The factors $\hat{P}^{(0)}$ are the regularised splitting functions. We refer the reader to section 4 of ref.~\cite{Bizon:2017rah} for the definition of the luminosity factors and their ingredients.
We further defined $\zeta_{si} \equiv k_{tsi}/k_{t1}$ and introduced the notation $\dZ$ to denote an ensemble that describes the emission of $n$ identical independent blocks.
In this notation, we define the average of a function $G(\Phi_B,\{k_i\})$ over the measure
$\rd {\cal Z}$  as ($\zeta_{i} \equiv k_{ti}/k_{t1}$),
\begin{equation}
\label{eq:dZ}
\begin{split}
\int \! \!    \dZ  G(\Phi_B,\{k_i\})\! =\! \re^{-{\tildeR}'(k_{t1})\ln\frac{1}{\eps}} \! 
   \sum_{n=1}^{\infty}\!  \frac{1}{n!} \! \prod_{i=2}^{n+1} \! 
    \int_{\eps}^{1} \!  \! \frac{\rd\zeta_i}{\zeta_i} \!  \!  \int_0^{2\pi}\!  \! 
   \frac{\rd\phi_i}{2\pi} \!  {\tildeR}'(k_{t1})G(\Phi_B,k_1,\dots,k_{n+1})\,.
\end{split}
\end{equation}
We stress that the $\ln 1/\eps$ divergence that appears in the exponential prefactor of eq.~\eqref{eq:dZ} cancels exactly against that contained in the resolved real radiation, which is instead encoded in the nested sums of products on the right-hand side of the same equation.

To obtain eq.~\eqref{eq:master-kt-space}, we exploited the rIRC safety of the observable to expand all the ingredients in eq.~\eqref{eq:master-kt-space} about $k_{t1}$ since $\zeta_i =  k_{ti}/k_{t1} \sim \mathcal O(1)$.
Indeed, rIRC safety guarantees that blocks with $k_{ti} \ll k_{t1}$ are fully cancelled by the term 
$\exp\{-{\tildeR}'(k_{t1})\ln(1/\eps)\}$ of eq.~\eqref{eq:dZ}.
Such an expansion is not strictly necessary, but makes a numerical implementation much more efficient.
Because eq.~\eqref{eq:master-kt-space} was expanded about $k_{t1}$, it contains explicitly the derivatives
\begin{equation}
{\tildeR}'= \rd \tildeR/\rd\tildeL\,,\qquad {\tildeR}''= \rd
      {\tildeR}'/\rd\tildeL\,,\qquad {\tildeR}'''= \rd {\tildeR}''/\rd\tildeL
\end{equation}
of the radiator $\tildeR$, which is given by
\begin{align}
\label{eq:mod-radiator}
\tildeR(k_{t1}) &= - \tildeL g_1(\as \beta_0\tildeL ) -
  g_2(\as \beta_0\tildeL ) - \frac{\as}{\pi}
  g_3(\as \beta_0\tildeL ) - \frac{\as^2}{\pi^2}
  g_4(\as \beta_0\tildeL )\,,
\end{align}
with $\as = \as(\mu_R)$ and $\mu_R\sim M$ being the renormalisation scale.
The functions $g_i$ are reported in eqs.~(B.8-B.11) of ref.~\cite{Bizon:2017rah}. 
With this choice, the logarithmic accuracy is effectively defined in terms of $L=\ln({Q}/{k_{t1}})$, where $Q \sim M$ is the resummation scale, whose variation is used to probe the size of missing logarithmic higher-order corrections in eq.~\eqref{eq:master-kt-space}.
We refer the reader to ref.~\cite{Bizon:2017rah} for further details.
 
Though eq.~\eqref{eq:master-kt-space} is valid for inclusive observables, the \textsc{RadISH} formalism can be systematically extended to any transverse observable in colour-singlet production, as stressed in ref.~\cite{Bizon:2017rah}. 
It is particularly instructive to consider the case of jet-veto resummation, which is currently available at NNLL~\cite{Becher:2012qa,Banfi:2012jm,Becher:2013xia,Stewart:2013faa,Banfi:2013eda,Michel:2018hui}. 
At this logarithmic accuracy the resummation must include an additional \textit{clustering} correction since
the jet algorithm can cluster two independent emissions close in the pseudo-rapidity $\eta$ or in the azimuthal angle $\phi$. Moreover, the resummation must  account for a \textit{correlated} correction, which amends the inclusive treatment of the correlated squared amplitude for two emissions~\cite{Dokshitzer:1997iz}, accounting for configurations where the two correlated emissions are not clustered in the same jet~\cite{Banfi:2012jm}.
The analytical result for the NNLL resummation reads~\cite{Banfi:2012jm}
\begin{align}\label{eq:jetvetoanalytic}
\frac{\rd\sigma(\ptj)}{\rd\Phi_B} ={\tildecalL}_{\rm
  NNLL}(\ptj) (1 + \mathcal F_{\rm clust} +  \mathcal F_{\rm corr} ) e^{\tildeL g_1(\as \beta_0\tildeL ) +
  g_2(\as \beta_0\tildeL ) + \frac{\as}{\pi}
  g_3(\as \beta_0\tildeL ) }\,,
\end{align}
where the functions $g_1$, $g_2$ and $g_3$ are the same appearing in the radiator of eq.~\eqref{eq:mod-radiator} for the colour singlet $\pt$, and now $L = \ln(Q/\ptj)$.
For generalised $k_t$ algorithms~\cite{Catani:1993hr,Ellis:1993tq,Dokshitzer:1997in,Wobisch:1998wt,Cacciari:2008gp} the clustering and the correlated corrections at NNLL accuracy in the limit of small jet radius $R$ read\footnote{For the full formul\ae\ see the appendix of ref.~\cite{Banfi:2012jm}.}
\begin{align}
	\mathcal F_{\rm clust} &= \frac{4 C \as^2  (\ptj) }{\pi^2} L \left(- \frac{\pi^2 R^2}{12} + \frac{R^4}{16}\right)\,,\\
	\mathcal F_{\rm corr} &= \frac{4 C \as^2  (\ptj) }{\pi^2} L  \left[ \left(\frac{(-131+12\pi^2 + 132 \ln 2) C_A}{72} + \frac{(23 -24 \ln 2) n_f}{72}\right)\ln \frac{1}{R} \right. \notag \\
	& \left. \phantom{\ln \frac{1}{R}} \qquad \qquad \quad +0.61 C_A -0.015 n_f \right]+ \mathcal O(R^2)\,,
\end{align}
 where $C$ is  $C_F = 4/3$ or $ C_A =3 $ for incoming quarks or incoming gluons, respectively.
 The inclusion of small-$R$ resummation at LL$_R$ accuracy was found to have, for gluon-initiated processes, a very moderate effect on the result for values of $R$ typically used in phenomenological applications~\cite{Banfi:2015pju}.
 In this work we neglect these effects, under the assumption that their impact is negligible. Further studies in this 
 direction are beyond the scope of this paper.

We can now recast the resummation for $\ptj$ in the \textsc{RadISH} language as
\begin{align}\label{eq:jetvetoradish}
\frac{\rd\sigma(\ptj)}{\rd\Phi_B}=  \frac{\rd\sigma^{\rm incl}(\ptj)}{\rd\Phi_B} +  \frac{\rd\sigma^{\rm clust}(\ptj)}{\rd\Phi_B} +\frac{\rd\sigma^{\rm corr}(\ptj)}{\rd\Phi_B}\,,
\end{align}
where
\begin{align}
 & \frac{\rd\sigma^{\rm incl}(\ptj)}{\rd\Phi_B} = \int\frac{\rd k_{t1}}{k_{t1}}\frac{\rd
  \phi_1}{2\pi} \,d{\cal Z} \,\partial_\tildeL\left[-\re^{-\tildeR_{\rm NNLL}(\kto)}{\tildecalL_{\rm NNLL} (\kto)}
    \right]\Theta\Big(\ptj-\kto\Big)\,,\\ \notag\\
&\frac{\rd\sigma^{\rm clust}(\ptj)}{\rd\Phi_B} = \int\frac{\rd k_{t1}}{k_{t1}}\frac{\rd
  \phi_1}{2\pi} \,d{\cal Z} \, \re^{-\tildeR(\kto)}{\cal \tildeL_{\rm
  NLL} }(\kto)\tildeR'(\kto)
  \int d\Delta y_{1 s_1}\frac{d\Delta\phi_{1
  s_1}}{2\pi}\int^{\kto}_0\frac{ dk_{ts_1}}{k_{ts_1}} \,
 \notag\\
& \times\!  \left(2 C_{\ell_{s_1}} \frac{\as(\kto)}{\pi} \right)  J_R(k_1,k_{s_1})
      \!\bigg[
     \Theta\Big(\ptj-|\vec{k}_{t1} + \vec{k}_{ts_1}|\Big)\! -\! \Theta\Big(\ptj-\kto\Big) 
     \bigg]\Theta\Big(\ptj-\max_{i > 1}\{k_{t,i}\}\Big)+\notag\\
&\frac{1}{2!}\! \int \! \frac{\rd k_{t1}}{k_{t1}}\frac{\rd
  \phi_1}{2\pi}d{\cal Z} \re^{- \tildeR(\kto)}{\cal \tildeL_{\rm
  NLL}}(\kto)\left(\tildeR'(\kto)\right)^2 \!  \! \! 
  \int \!  d\Delta y_{s_1 s_2}\frac{d\Delta\phi_{s_1
  s_2}}{2\pi} \!  \!  \! \int^{\kto}_0 \!  \!  \frac{ dk_{ts_1}}{k_{ts_1}}\frac{\rd
  \phi_{s_1}}{2\pi} \!  \!  \int^{\kto}_0\frac{ dk_{ts_2}}{k_{ts_2}} \notag\\
& \left(\!2 C_{\ell_{s_2}} \! \frac{\as(\kto)}{\pi}\! \right)  \! \! J_R(k_{s_1},k_{s_2})
      \!\bigg[
     \Theta\Big(\ptj-|\vec{k}_{ts_1} + \vec{k}_{ts_2}|\Big)\! - \!\Theta\Big(\ptj-\max\{k_{ts_1},k_{ts_2}\}\Big) 
     \bigg] \! \Theta\Big(\ptj-\kto \!\Big)\,,\\ \notag  \\
& \frac{\rd\sigma^{\rm corr}(\ptj)}{\rd\Phi_B} = \int\frac{\rd k_{t1}}{k_{t1}}\frac{\rd
  \phi_1}{2\pi} \,d{\cal Z} \, \re^{-\tildeR(\kto)}{\cal \tildeL_{\rm
  NLL}} (\kto)\tildeR'(\kto)
  \int d\Delta y_{1 s_1}\frac{d\Delta\phi_{1
  s_1}}{2\pi}\int^{\kto}_0\frac{ dk_{ts_1}}{k_{ts_1}} \, \notag\\
  & \times \left(2 C_{\ell_{s_1}} \frac{\as(\kto)}{\pi} \right)  \times \mathcal C\left(\Delta y_{1 s_1},\Delta \phi_{1 s_1},\frac{\kto}{k_{ts_1}}\right) (1-J_R(k_1,k_{s_1})) \notag \\
  & \times  \!\bigg[\Theta\Big(\ptj-\kto\Big) -
     \Theta\Big(\ptj-|\vec{k}_{t1} + \vec{k}_{ts_1}|\Big)
     \bigg]\Theta\Big(\ptj-\max_{i > 1}\{k_{t,i}\}\Big)+ \notag\\
  & \frac{1}{2!}\! \int \! \frac{\rd k_{t1}}{k_{t1}}\frac{\rd
  \phi_1}{2\pi}d{\cal Z} \re^{- \tildeR(\kto)}{\cal \tildeL_{\rm
  NLL}}(\kto)\left(\tildeR'(\kto)\right)^2 \!  \! \! 
  \int \!  d\Delta y_{s_1 s_2}\frac{d\Delta\phi_{s_1
  s_2}}{2\pi} \!  \!  \! \int^{\kto}_0 \!  \!  \frac{ dk_{ts_1}}{k_{ts_1}}\frac{\rd
  \phi_{s_1}}{2\pi} \!  \!  \int^{\kto}_0\frac{ dk_{ts_2}}{k_{ts_2}} \notag\\
  & \times \left(2 C_{\ell_{s_2}} \frac{\as(\kto)}{\pi}\right) \mathcal C\left(\Delta y_{s_1 s_2},\Delta \phi_{s_1 s_2},\frac{k_{ts_2}}{k_{ts_1}}\right) (1-J_R(k_{s_1},k_{s_2})) \notag \\
  & \times   \!\bigg[\Theta\Big(\ptj-\max\{k_{ts_1},k_{ts_2}\}\Big) -\Theta\Big(\ptj-|\vec{k}_{ts_1} + \vec{k}_{ts_2}|\Big)
     \bigg]\Theta\Big(\ptj-\kto\Big)\,.
\label{eq:sigma-NNLL-ptjv}
\end{align}
Here $J_R(k_a,k_b) = \Theta(R^2 - (\Delta y_{ab})^2 - (\Delta \phi_{ab})^2)$, where $\Delta y_{ab}$ and $\Delta \phi_{ab}$ are the difference in rapidity and in azimuthal angle between two emissions $a$ and $b$, and $C_{\ell}$ is the colour factor associated with the incoming hard leg $\ell$.
The function $\mathcal C$ is defined as the ratio of the correlated part of the double-soft squared amplitude and the product of the two single-soft squared amplitudes, cfr. eq.~(30) of ref.~\cite{Monni:2019yyr}.

At NNLL accuracy the formulations of eqs.~(\ref{eq:jetvetoanalytic}) and (\ref{eq:jetvetoradish}) are equivalent, the only difference being in the treatment of subleading terms. 
The jet veto resummation of eq.~(\ref{eq:jetvetoanalytic}) has the advantage that it is fully analytic, thereby allowing for a simple and fast implementation in a numerical code.
The \textsc{RadISH} formulation, on the other hand, features less compact formul\ae\ which need to be evaluated numerically. For the single-differential jet-veto resummation we employ the implementation 
of the analytic formul\ae\ in eq.~\eqref{eq:jetvetoanalytic} since their evaluation is faster. By casting the jet-veto resummation in the \textsc{RadISH} formulation, however, one gains a more differential description of the radiation than the one provided by eq.~\eqref{eq:jetvetoanalytic}.
This fact can be exploited to formulate the joint resummation of logarithms of $p_T$ and $\ptj$ within this framework by noticing that the two observables share the same Sudakov radiator $\tildeR_{\rm NNLL}$~\cite{Banfi:2012jm} at NNLL accuracy. The double-differential resummation is then achieved by supplementing the phase space constraint for the inclusive $\pt$ resummation of eq.~\eqref{eq:master-kt-space} with a veto requirement, and by adding the clustering and correlated corrections that we described above. 
The interested reader can find the resulting formul\ae\ in the supplementary material of ref.~\cite{Monni:2019yyr}.

The \textsc{RadISH} code implements the above formul\ae\ for the resummation of transverse observables using Monte Carlo methods.
We refer the reader to section~4.3 of ref.~\cite{Bizon:2017rah} for details of the Monte Carlo evaluation and on event generation in the \textsc{RadISH} code.

\subsection{Matching of resummation and fixed-order predictions}\label{sec:matching}

The calculation of a physical prediction for a general observable $v$ across its entire differential spectrum requires a consistent matching 
between the fixed-order distribution valid in the hard region (large $v$) and the resummed result valid in the soft/collinear region (small $v$).
This implies that, on the one hand, resummation effects have to vanish at large $v$, while, on the other hand, 
the fixed-order contribution should vanish at small $v$.
In order to suppress resummation effects at large $v$, we map the limit $\kto \rightarrow Q$, where the logarithms vanish, onto $\kto \rightarrow \infty$ by introducing modified logarithms
\begin{equation}\label{eq:modlog}
	\ln \frac{Q}{\kto} \rightarrow \tilde{L} \equiv\frac{1}{p} \ln \left( \left(\frac{Q}{\kto}\right)^p + 1 \right),
\end{equation}
where $p$ is a positive real parameter.
Its value is chosen such that the resummed component decreases faster than the fixed-order spectrum for $v \gtrsim 1$.
As a consequence, the logarithms in the Sudakov radiator~\eqref{eq:mod-radiator}, its derivatives and the luminosity factors have to be replaced by $\tilde L$.
For consistency,
eqs.~\eqref{eq:master-kt-space} and \eqref{eq:jetvetoradish} are supplemented by the following Jacobian in accordance with the 
replacement in eq.~\eqref{eq:modlog}:
\begin{equation}
	\mathcal{J} (\kto) =  \left(\frac{Q}{\kto}\right)^p \left( \left(\frac{Q}{\kto}\right)^p + 1 \right)^{-1}.
\end{equation}
This prescription leaves the $\Theta$ functions in eq.~\eqref{eq:master-kt-space} and eq.~\eqref{eq:jetvetoradish} unchanged 
and modifies the final result by introducing power corrections in $(Q/\kto)^p$ beyond the nominal accuracy.

In order to perform the matching of the resummed calculation in the \textsc{RadISH} formalism 
to the fixed-order prediction, it is useful to introduce the cumulative fixed-order cross section, since the resummation 
is defined at this level,
\begin{align}\label{eq:cumulative-NNLO}
	\sigma_{\rm NNLO}^F (v) = \int_0^{v} \rd v' \, \frac{\rd \sigma_{\rm NNLO}^F(v')}{\rd v'} = \sigma_{\rm NNLO}^F - \int_{v}^\infty \rd v' \,\frac{\rd \sigma_{\rm NLO}^{F+\text{jet}}(v')}{\rd v'}\,,
\end{align}
where $\rd \sigma_{\rm NNLO}^F$ is the fully differential NNLO cross section of eq.~\eqref{eq:qtsub}, and $\sigma_{\rm NNLO}^F$ is the NNLO cross section integrated over radiation.
Note that in all expressions throughout this section we have dropped the explicit dependence on the Born phase space $\Phi_B$ for the sake of brevity, which shall be understood implicitly for all cross sections. 
Thus, the cumulative NNLO cross section in eq.~\eqref{eq:cumulative-NNLO} is fully differential in 
the Born kinematics such that arbitrary fiducial cuts on the colour-singlet final state can be applied. 
We stress that the integral from 0 to $v$ of $\rd \sigma^F_{\rm NNLO}$ is well-defined since all IR divergences have been canceled through the $q_T$ subtraction procedure.
For brevity, we also drop the superscript $F$ referring to a general colour-singlet final state from 
the NNLO cross section in the following equations.

We recall that the NNLO accuracy as denoted in eq.~\eqref{eq:cumulative-NNLO} applies to cumulative (integrated) cross sections, whereas
the differential spectrum at large $v$ is only NLO accurate.
In the next section we will show results at the cumulative and at the differential level; in both cases, we shall label the 
fixed-order accuracy according to the accuracy defined at the cumulative level.\footnote{Note that this convention is different from that used in ref.~\cite{Bizon:2017rah,Bizon:2018foh,Bizon:2019zgf}.}

There is some freedom when defining a procedure to match resummation
and fixed-order predictions:
At a given perturbative order various schemes can be defined that differ from one another only by subleading terms, i.e. 
beyond the formal accuracy of the calculation. 
In the \textsc{Matrix+RadISH} interface we offer the possibility to choose between two different matching schemes 
to assess the associated uncertainties (see appendix~\ref{sec:parameters-settings}).
The first scheme is a customary additive scheme, which at NNLO+NNLL is defined as
\begin{align}\label{eq:additive}
{\rm \sigma}^{\rm add.\, match.}_{\rm NNLO+NNLL}(v)  = \sigma_{\rm NNLO}(v) - \left[\sigma_{\rm NNLL}(v)\right]_{\rm NNLO} + \sigma_{\rm NNLL }(v)\,.
\end{align} 
Here $ [\ldots]_{{\rm N}^k{\rm LO}} $ indicates the expansion of the expression inside the bracket truncated at N$^k$LO,
such that the second term is the expansion of the resummed cross section $\sigma_{\rm NNLL }(v)$ up to NNLO (i.e. ${\cal O}(\as^2)$).
It subtracts the logarithmically enhanced contributions at small $v$ from the fixed-order component, thereby turning it finite and avoiding a 
double counting between the first and the third term. 
The second scheme is a multiplicative scheme,
\begin{equation}
\label{eq:multiplicative1}
{\rm \sigma}^{\rm mult. \,match.}_{\rm NNLO+NNLL}(v) = \frac{{\rm \sigma}_{\rm NNLL}(v)}{{\rm \sigma}_{\rm NNLL}^{\rm asym.} } \left[{\rm \sigma}_{\rm NNLL}^{\rm asym.} \frac{{\rm \sigma}_{\rm NNLO}(v)}{\left[\sigma_{\rm NNLL}(v)\right]_{\rm NNLO}}\right]_{\rm NNLO},
\end{equation}
as formulated in refs.~\cite{Caola:2018zye,Bizon:2018foh}.
The quantity $ {\rm \sigma}_{\rm NNLL}^{\rm asym.}$ is defined as the asymptotic ($v \gg 1$) limit of the resummed cross section
\begin{equation}
{\rm \sigma}_{\rm NNLL}(v)\xrightarrow[v \gg 1 ]{} {\rm \sigma}_{\rm NNLL}^{\rm asym.}\,.
\end{equation}
This prescription ensures that in the limit $v \to 0$ eq.~\eqref{eq:multiplicative1} reduces to the resummed prediction, and that for 
$v \gg 1$  it reproduces the fixed-order result.
The main difference between the two procedures is that the multiplicative approach is more robust against 
numerical instabilities at very low $v$, since a potential miscancellation of the logarithmic terms between the NNLO result 
and the expansion is suppressed by the resummation factor ${\rm \sigma}^{\rm NNLL}(v)$.
Analogous formul\ae\ can be derived at NNLO+N$^3$LL and at NLO+NLL accuracy.
The detailed matching formul\ae\ for the multiplicative matching scheme 
are reported in
appendix~A of ref.~\cite{Bizon:2018foh}.
We finally note that in both matching schemes the cumulative cross section at $v\to\infty$ tends to $\sigma_{\rm NNLO}$ and by construction the differential distribution fulfils the unitarity constraint, i.e.\ its integral yields the NNLO cross section.

The matching procedures immediately generalise to the double-differential case by rewriting the equations with the double-cumulative cross section $\sigma (v_1, v_2)$, obtained by integrating over the double-differential distribution ${\rd \sigma(v_1,v_2)}/{\rd v_1 \rd v_2}$.
In particular, the multiplicative scheme used for the double-differential case is
\begin{equation}\label{eq:multiplicative2}
{\rm \sigma}^{\rm mult. \,match.}_{\rm NNLO+NNLL}(v_1,v_2) = \frac{{\rm \sigma}_{\rm NNLL}(v_1,v_2)}{{\rm \sigma}_{\rm NNLL}^{\rm asym.} } \left[{\rm \sigma}_{\rm NNLL}^{\rm asym.} \frac{{\rm \sigma}_{\rm NNLO}(v_1, v_2)}{\left[\sigma_{\rm NNLL}(v_1,v_2)\right]_{\rm NNLO}}\right]_{\rm NNLO}\,.
\end{equation}
We observe that by using the multiplicative scheme \eqref{eq:multiplicative2} one automatically recovers the NNLO+NNLL result \eqref{eq:multiplicative1} for $v_1$ ($v_2$) in the limit $v_2 \rightarrow \infty$ ($v_1 \rightarrow \infty)$.
Unless explicitly stated otherwise, multiplicative matching is used for all results presented in this paper.
We recall that the matched cross sections in eqs.~\eqref{eq:additive}, \eqref{eq:multiplicative1} and~\eqref{eq:multiplicative2} shall 
be understood as being fully differential in the Born level momenta, allowing us to 
apply arbitrary fiducial cuts on the colour-singlet final states.

The multiplicative scheme has one further advantage: when matching the resummation at NNLL to the 
fixed-order prediction at NNLO using eq.~\eqref{eq:multiplicative1} or eq.~\eqref{eq:multiplicative2}, the terms that are constant in $v$ 
at $\mathcal{O}(\as^2)$ are automatically included in the resummation, since the fixed-order contribution 
multiplies the resummed cross section.
This procedure correctly resums the whole tower of $\as^n L^{2n-4}$ contributions, which are formally part of the N$^3$LL correction, in the matched cross section.
Although all our NNLO+NNLL predictions include these additional corrections,
we refrain from explicitly adopting a specific notation in the remainder of this paper and assume it to be understood.
In particular, the constant terms at $\mathcal{O}(\as^2)$ contain the two-loop virtual corrections and the second-order collinear coefficient functions.
In an additive matching, on the other hand, the terms $\as^n L^{2n-4}$ are included only up to $n=2$ through the fixed-order contribution, so that
they simply amount to a constant shift in the matched cross section at the cumulative level. Since the constant terms are still unknown analytically
for jet-veto resummation and the joint resummation of $\pt$ and $\ptj$, if we were to use the additive matching scheme, the additional logarithmic corrections would not be included.

\subsection{Numerical implementation and validation}\label{sec:implementation}

In the following, we briefly discuss some details of the practical implementation and the numerical validation of the {\sc Matrix+RadISH} results.
All fixed-order ingredients are evaluated by \textsc{Matrix}, which provides the cumulative distributions at NLO and at NNLO.
\textsc{Matrix} also generates the Born level phase space used to integrate the resummed component, and for a given kinematics it 
computes the process-dependent Born matrix element as well as the hard-virtual corrections at NLO and at NNLO.
For each Born event \textsc{RadISH} then produces the initial-state radiation using the numerical algorithm 
described in section~\ref{sec:radish} to perform the resummation of large logarithmic contributions.
In parallel, \textsc{RadISH} also evaluates the expansion and the asymptotic limit of the resummed cross section entering eqs.~\eqref{eq:additive} and \eqref{eq:multiplicative1}.\footnote{We refer the reader to section~4.2 of ref.~\cite{Bizon:2017rah} for details on the numerical evaluation of the terms entering in the expansion.}
Furthermore, we perform on-the-fly variations of the renormalisation, the factorisation and the resummation scale
for each Born event. After all ingredients have been integrated to a sufficient numerical precision, the matching 
is performed according to either eq.~\eqref{eq:additive} or eq.~\eqref{eq:multiplicative1} as a post-processing 
step of the calculation.

Our calculations have been extensively validated by performing a variety of tests:
for \mbox{$2 \rightarrow 1$} processes, we have compared our resummed results against those obtained independently with the standalone version of the \textsc{RadISH} code where the required perturbative ingredients for Higgs boson and Drell--Yan production are implemented, finding
full agreement up to numerical uncertainties for the resummed predictions.
We have further checked for these processes that we find also full agreement at the level of matched cross sections calculated with \textsc{Matrix+RadISH} and those obtained by matching \textsc{RadISH} distributions with fixed-order predictions from \textsc{MCFM}~\cite{Campbell:2015qma,Boughezal:2016wmq}.

A very powerful check of the robustness of our predictions for more complex colour-singlet processes can be achieved by comparing the expansion of the resummed cross section to the fixed-order result in the small-$v$ limit, which we have performed for several 
$2\to 1$ and $2\to 2$ processes.
At NLL\,(NNLL) accuracy, the resummation predicts the logarithmically enhanced contributions appearing in the differential fixed-order distribution at small $v$ up to order $\as$\,($\as^2$).
The constant terms at the level of the NLO (NNLO) cumulative cross section, on the other hand, 
are not contained in the expansion up to $\as$\,($\as^2$) of the NLL (NNLL) cross section, such 
that $\sigma_{\rm (N)NLO}$ and $\left[\sigma_{\rm (N)NLL}\right]_{\rm (N)NLO}$ differ by a constant in the $v\to 0$ limit. The correct constant terms are included only in the resummed expression at the next
logarithmic order, i.e. the difference
$\sigma_{\rm NLO}$ and $\left[\sigma_{\rm NNLL}\right]_{\rm NLO}$ as well as $\sigma_{\rm NNLO}$ and $\left[\sigma_{\rm N^{3}LL}\right]_{\rm NNLO}$ tend to zero in the small-$v$ limit.
 
 \begin{figure}[t]
  \centering\hspace{-0.3cm}
  \includegraphics[trim={0 -0.2cm 0
    0},width=0.5\linewidth]{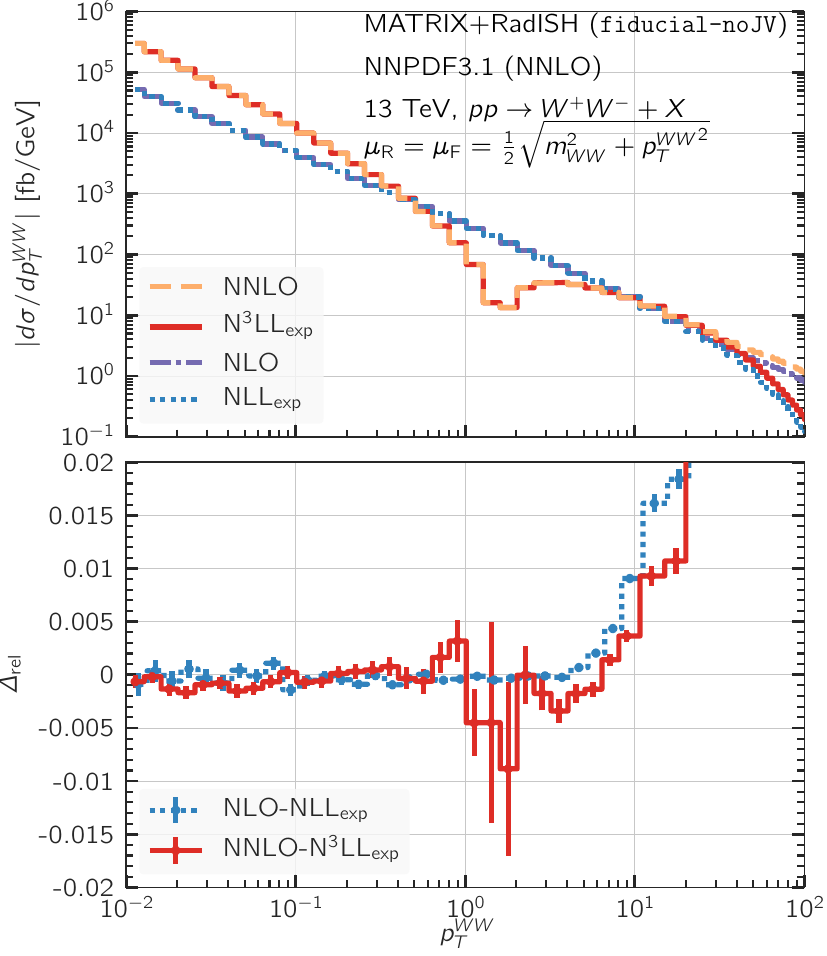} 
  \includegraphics[trim={0 -0.2cm 0
    0},width=0.485\linewidth]{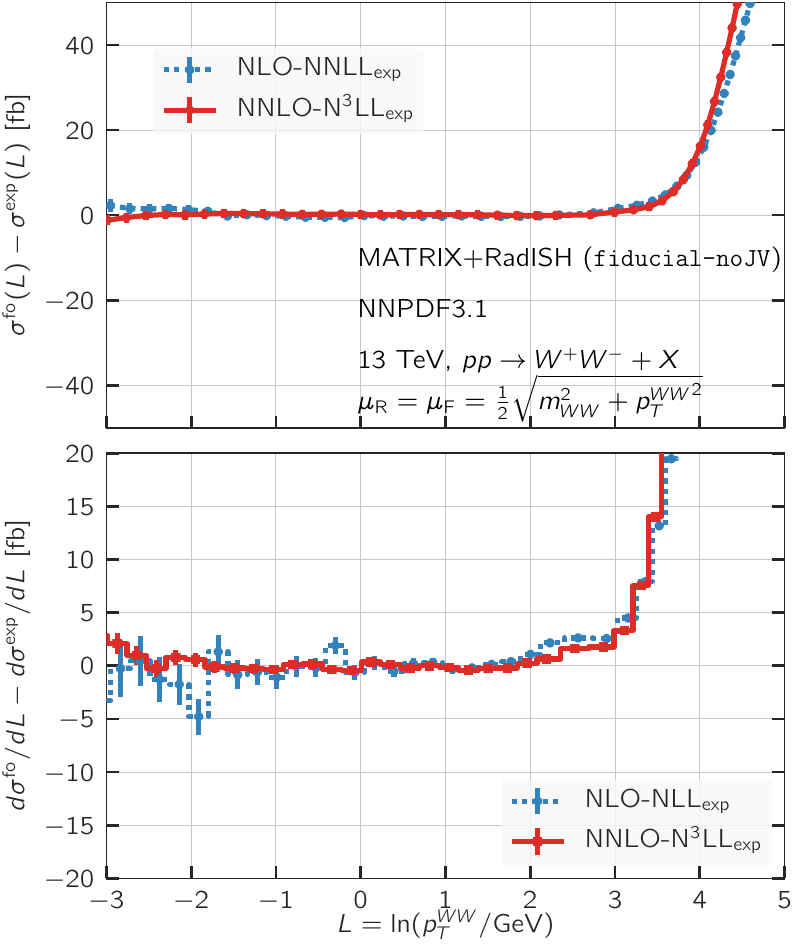} 
  \caption{Left plot: transverse-momentum spectrum of the $W^+ W^-$ pair 
  at NLO (purple, dot-dashed) and NNLO (orange, dashed), and the expansion of the NLL (blue, dotted) and N$^3$LL (red, solid) cross section. 
  The lower panel shows the relative difference eq.~\eqref{eq:reldiff} between the fixed-order cross section and the expansion.
  Right plot: the upper panel shows the difference at the cumulative level between NLO and NLL expansion (blue, dotted), and between NNLO and N$^3$LL expansion (red, solid). The lower panel shows the same results for the derivative of the cumulative cross 
  section with respect to $\ln(\ptWW/{\rm GeV})$.}
  \label{fig:expansion}
\end{figure}

Thus, in order to perform a non-trivial check on the validity of our calculation we consider $W^+W^-$ production
and study the small-$\ptWW$ behaviour of the differential $\ptWW$ distribution in the left plot of figure~\ref{fig:expansion}, 
where $\ptWW$ is the transverse momentum of the $W^+W^-$ pair. In the upper (lower) panel of the right plot of figure~\ref{fig:expansion} we show the 
corresponding cumulative (differential) cross section as a function of $\ln(\ptWW/{\rm GeV})$.
These results are obtained with the settings described in section~\ref{sec:setup} in the fiducial phase space 
defined as ``\texttt{fiducial-noJV}''.\footnote{Note that the dynamic scales and the fiducial cuts for the resummation and its expansion
are based on the Born level kinematics and thus correspond to those applied at fixed order only in the limit where extra QCD radiation is 
soft or collinear, and in particular in the $\ptWW\rightarrow 0$ limit.}
By using dedicated high-statistics runs we were able to push the comparison down to remarkably low values of $\ptWW = 0.05$\,GeV
with statistical uncertainties below the permille level.
The right plot of figure~\ref{fig:expansion} indicates an excellent agreement for the $\ptWW$ distribution 
between the fixed-order prediction and the respective logarithmic terms both at NLO and at NNLO.\footnote{We remind the reader that, when differential distributions are shown, the label NLO and NNLO refers to the $\mathcal{O}(\as)$ and the $\mathcal{O}(\as^2)$ prediction, respectively.}
The dip around 1-2 GeV is due to the fact that the NNLO spectrum and the NNLL expansion become negative. 
In particular the relative difference
\begin{equation}\label{eq:reldiff}
 	\Delta_{\rm rel} \equiv \left( \frac{{\rm d} \sigma^{\rm (N)NLO}}{{\rm d} \ptWW } -   \left[\frac{{\rm d} \sigma^{\rm (N)NLL}}{{\rm d} \ptWW }\right]_{\rm (N)NLO}\right)\Bigg/\left[\frac{{\rm d} \sigma^{\rm (N)NLL}}{{\rm d} \ptWW }\right]_{\rm (N)NLO}
\end{equation}
in the lower panel shows the striking cancellation between those terms at the few-permille level for very small $\ptWW$.
At the cumulative level, on the other hand, the upper frame of the right plot of figure~\ref{fig:expansion} nicely confirms
that the NLO$-$NNLL$_{\rm exp}$ and the NNLO$-$N$^3$LL$_{\rm exp}$ differences tend to zero, which validates our implementation of the constant terms at $\mathcal O(\as)$ and $\mathcal O(\as^2)$, respectively.  
Finally,  in the lower panel of that figure we show the absolute differences of the differential distributions by taking the derivatives 
with respect to $\ln{\ptWW}$. Although the information might appear slightly redundant, by considering the absolute difference here
we unambiguously prove that there are no potential residual differences also for the subleading logarithmic contributions.

We have performed similar checks for other $2\to1$ and $2\to 2$ colour-singlet processes, with various choices of the renormalisation scale $\mu_R$, the factorisation scale $\mu_F$, and the resummation scale $Q$.
In particular, we performed the exact same study for fully inclusive $W^+W^-$ production,
i.e.\ without fiducial cuts. In all cases we found an excellent agreement between the expansion of the resummed cross section and the fixed order result.


\section{Resummed predictions for $W^+ W^-$ production at the LHC}\label{sec:WWprod}

As a first application of the \textsc{Matrix+RadISH} framework introduced in section~\ref{sec:framework} we consider $W^+W^-$ production at the LHC.
This process plays a prominent role in the vast physics programme at the LHC since it has the largest cross section among all diboson 
production processes. Experimental measurements of the $W^+W^-$ cross section at the Tevatron~\cite{Aaltonen:2009aa,Abazov:2011cb} and at the LHC~\cite{ATLAS:2012mec,Chatrchyan:2013yaa,ATLAS:2014xea,Aad:2016wpd,Chatrchyan:2013oev,Khachatryan:2015sga,Aaboud:2017qkn,CMS:2016vww,Aaboud:2019nkz} 
provide crucial tests of the EW gauge sector of the SM and of the mechanism of EW symmetry breaking. Moreover, the production of $W$-boson pairs is an important probe of BSM physics.
Since the dynamics of $W^+ W^-$ production is sensitive to the value of the trilinear gauge coupling already at the Born level, 
$W^+ W^-$ measurements put stringent bounds on the strength of anomalous trilinear gauge couplings in indirect searches for new physics~\cite{ATLAS:2012mec,Chatrchyan:2013yaa,Khachatryan:2015sga,Aad:2016wpd}. 
$W^+ W^-$ production contributes also as irreducible background to direct searches for BSM particles decaying into leptons, missing energy and/or jets, 
and to Higgs boson measurements in the $H \rightarrow W^+ W^-$ decay channel.
Since the two neutrinos prevent the reconstruction of the full event kinematics, 
an accurate theoretical description of the $W^+ W^-$ final state is essential to 
enhance the experimental sensitivity in BSM searches and Higgs boson measurements.

Moreover, $W^+ W^-$ analyses generally apply a rather stringent cut on the jet activity in order to deplete the signal contamination due to top-quark backgrounds.
Such a jet veto introduces large logarithmic contributions in the calculation of the fiducial cross section, which challenges the validity of fixed-order predictions 
and induces an additional uncertainty in the extrapolation to the inclusive $W^+ W^-$ cross section.

As a consequence, $W^+ W^-$ production has received much attention in the past years in a joint effort to reduce theoretical uncertainties in order to match the increasing precision of the experimental data.
Next-to-leading order (NLO) QCD \cite{Ohnemus:1991kk,Frixione:1993yp} predictions for on-shell $W$ bosons have been known for years, 
and leptonic $W$ boson decays were included in refs.~\cite{Campbell:1999ah,Dixon:1999di,Dixon:1998py,Campbell:2011bn}.
Due to large $\mathcal O(\as)$ corrections, a substantial effort has been put into calculating contributions at $\mathcal O(\as^2)$.
The simplest $\mathcal O(\as^2)$ correction is the loop-induced $gg \rightarrow W^+ W^-$ subprocess, which is enhanced by the gluon luminosities.
Leading order (LO) predictions for the loop-induced gluon fusion channel were studied in refs.~\cite{Dicus:1987dj,Glover:1988fe,Binoth:2005ua,Binoth:2006mf,Campbell:2011bn,Campbell:2011cu}.
The full NNLO QCD corrections were calculated first for the inclusive $W^+ W^-$ cross section in the on-shell approximation in ref.~\cite{Gehrmann:2014fva}, 
and were later advanced to the fully differential level including leptonic decays in ref.~\cite{Grazzini:2016ctr}.
Contrary to the common lore, the quark-initiated contributions were found to be the dominant NNLO QCD corrections at the inclusive level, with the loop-induced $gg$-initiated channel
contributing only $\sim 30 \%$ of the full $\mathcal O(\as^2)$ correction. Various efforts have been made to go beyond NNLO QCD accuracy:
NLO EW corrections were calculated for stable $W$ bosons~\cite{Bierweiler:2012kw,Baglio:2013toa,Billoni:2013aba}, and with a consistent treatment of the leptonic decays~\cite{Biedermann:2016guo,Kallweit:2019zez}.
The presumably dominant $\mathcal O(\as^3)$ corrections are known from the calculation of the loop-induced gluon fusion 
contribution at NLO QCD~\cite{Caola:2015rqy,Caola:2016trd,Grazzini:2020stb}.
The first combination of NNLO QCD predictions with NLO EW corrections was presented very recently in ref.~\cite{Kallweit:2019zez}, 
and ref.~\cite{Grazzini:2020stb} performed their further combination with NLO QCD corrections to the loop-induced $gg$ channel, 
yielding the best fixed-order prediction for the $W^+ W^-$ process to date.

Analytic resummation approaches for different observables have been subject to various studies of $W^+W^-$ production. 
The transverse-momentum spectrum of on-shell $W^+ W^- $ pairs was calculated at NNLO+NNLL accuracy in ref.~\cite{Grazzini:2015wpa}, and the resummation of jet veto logarithms at NNLO+NNLL accuracy was performed in ref.~\cite{Dawson:2016ysj}.
In particular, the proper modelling of jet-vetoed $W^+ W^- $ cross sections has attracted much attention in the theory community~\cite{Jaiswal:2014yba,Becher:2014aya,Monni:2014zra,Dawson:2016ysj}, which has shown that higher-order corrections in both the perturbative and the logarithmic expansion are crucial to obtain
reliable predictions and uncertainty estimates in presence of a jet veto cut in the fiducial phase space.

\subsection{Outline of the calculation}

We consider the process
\begin{align}
\label{eq:process}
pp\to \ell^+\ell^{\prime\, -}\nu_{\ell}{\bar\nu}_{\ell^\prime}+X, 
\end{align}
where the charged final-state leptons and the corresponding (anti-)neutrinos have different flavours ($\ell\neq\ell^\prime$). 
All resonant and non-resonant contributions to this process are accounted for, including off-shell effects and spin correlations,
employing the complex-mass scheme~\cite{Denner:2005fg} without any resonance approximation.
We have full control on the momenta of the final-state leptons, which allows us to evaluate resummed cross sections in presence
of fiducial selection criteria as employed by the experiments. 
Our calculation applies to an arbitrary combination of (massless) leptonic flavours,
$\ell,\ell^\prime\in \{e,\mu\}$, and in order to compare against
experimental data, we evaluate the process $pp\to \ell^+ \ell^{\prime -}\nu_\ell {\bar \nu}_{\ell^\prime}+X$ with $\ell,\ell^\prime=e$ or $\mu$ and $\ell\neq \ell^\prime$.
For the sake of simplicity, this process will be denoted as $W^+W^-$ production in the following.

\begin{figure}[t]
\begin{center}
\hspace*{-0.3cm}
\begin{tabular}{cccc}
\includegraphics[height=1.8cm]{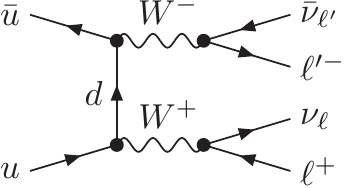} &
\includegraphics[height=1.8cm]{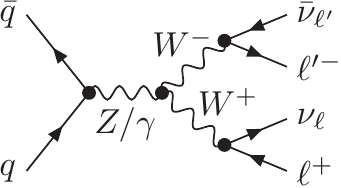} &
\includegraphics[height=1.8cm]{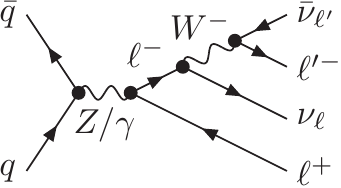} &
\includegraphics[height=1.8cm]{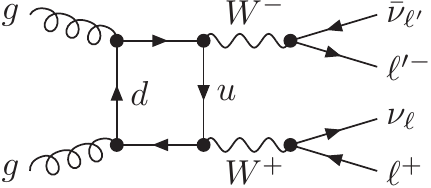} \\
(a) & (b) & (c) & (d)\\[0.3cm]
\end{tabular}
\caption[]{\label{fig:diag}{Feynman diagrams for $W^+W^-$ production: (a-c) sample tree-level diagrams in the quark-annihilation channel contributing at LO; (d) sample loop-induced diagram in the gluon fusion channel contributing at the NNLO.}}
\end{center}
\end{figure}

In figure~\ref{fig:diag}\,(a--c) we show representative Feynman diagrams at LO for $W^+W^-$ production. They are driven by quark annihilation in the initial state and include 
$t$-channel $W^+W^-$ topologies~(panel a), $s$-channel $W^+W^-$ topologies~(panel b), and $s$-channel Drell--Yan-type topologies~(panel c).
Figure~\ref{fig:diag}\,(d), on the other hand, shows a loop-induced diagram that is driven by gluons in the initial state, which 
enters the cross section at ${\cal O}(\as^2)$ and is part of the NNLO corrections. In an NNLO calculation the loop-induced gluon fusion 
contribution is, however, effectively only LO accurate and has Born kinematics. 
Without associated QCD radiation at fixed order, it contributes trivially to the differential observables we are considering in this paper.
At the resummed level, the LL resummation of this contribution starts at $\mathcal{O}(\as^3 L^2)$, and it is thus of the same size as N$^3$LL corrections to the 
$q \bar q$ channel. In principle the resummation of the loop-induced $gg$ channel
can be included alongside the resummation of the $q \bar q$ channel.\footnote{For jet-veto resummation this was done in ref.~\cite{Dawson:2016ysj}.}
However, in terms of their respective resummation both contributions
can be treated completely independently, and we refrain from considering the loop-induced gluon fusion contribution in what follows, since we reckon that its 
proper treatment requires to go beyond an effective accuracy of LO+LL.
In fact, NLO QCD corrections to the loop-induced  
gluon fusion contribution have already been evaluated within the \textsc{Matrix} framework for both $ZZ$~\cite{Grazzini:2018owa} and $W^+W^-$~\cite{Grazzini:2020stb} production.
They are strongly enhanced by the large gluon luminosity, and contributing at ${\cal O}(\as^3)$ they constitute the dominant N$^3$LO corrections to the cross sections of these two processes.
These corrections involve diagrams with real QCD radiation that would
directly contribute to the resummed observables under consideration. Therefore, a matching of the NLO QCD cross section with NNLL resummation  
for the loop-induced gluon fusion contribution would be a useful addition to the predictions presented below, 
which could be achieved within \textsc{Matrix+RadISH}, 
but it is beyond the scope of 
the present paper and left for future studies.

The calculation of higher-order corrections to the $W^+W^-$ production process in QCD 
perturbation theory is affected by a subtle interplay with contributions stemming 
from the production of off-shell top quarks, which mix through $t\rightarrow Wb$ decays~\cite{Dittmaier:2007th,Cascioli:2013gfa,Gehrmann:2014fva,Grazzini:2016ctr}
with real-radiation diagrams involving final-state bottom quarks. 
Due to the large cross section of top-quark processes at the LHC, these contributions induce a sizeable contamination of the $W^+W^-$ cross section. 
In order to deal with this problem, we employ the
four-flavour scheme (4FS), where bottom quarks are treated as massive and
do not appear as initial-state particles.
The bottom-quark mass renders partonic subprocesses with bottom quarks in the final state separately finite.
We then evaluate top-free $W^+W^-$ cross sections 
by omitting all subprocesses with real bottom-quark emissions,
which in turn are considered to be part of the (off-shell) top-pair background.
An alternative definition of the top-free $W^+W^-$ cross section can be obtained in the five-flavour scheme (5FS), where   
bottom quarks are treated as massless, by exploiting the scaling behaviour of the cross section with the top-quark width, which 
has been explained in detail in refs.~\cite{Gehrmann:2014fva,Grazzini:2016ctr}. 
As it has been shown there, the top-free $W^+W^-$ cross sections resulting from the 4FS and 5FS prescriptions 
agree within $1\%-2\%$, both at the inclusive level and with fiducial cuts.
Consequently, we exploit the simpler 4FS prescription throughout this paper.
We note that this approach requires the use of consistent sets of parton distribution functions (PDFs) with $n_f=4$ light parton flavours.

We recall that our implementation allows us to obtain resummed predictions for different transverse observables in $W^+W^-$ production at the LHC.
We calculate the transverse-momentum spectrum of $W^+W^-$ pairs at NNLO+N$^3$LL accuracy as well as the transverse-momentum 
spectrum of the leading jet (and equivalently the jet-vetoed cross section) at NNLO+NNLL accuracy. Even more notably, we also perform 
the simultaneous resummation of both observables at the double-differential level at NNLO+NNLL accuracy, which allows us to evaluate the 
spectrum of one of the two observables in presence of a veto on the other one.
For all differential observables presented here it is the first time that such accuracies are achieved for a nontrivial process, i.e.\ beyond \mbox{$2\to1$} scattering like Higgs boson production or the Drell--Yan process.
In addition, arbitrary fiducial selection criteria can be applied in phase space of 
the leptonic final states.

\subsection{Phenomenological results}\label{sec:WW}
\subsubsection{Setup}\label{sec:setup}

We present resummed predictions for various observables in $pp\to \ell^+ \ell^{\prime -}\nu_\ell {\bar \nu}_{\ell^\prime}+X$ production
with $\ell,\ell^\prime=e$ or $\mu$ and $\ell\neq \ell^\prime$ at the LHC with $\sqrt{s}=13$\,TeV.
We employ the the $G_\mu$ scheme to evaluate the EW coupling 
$\alpha=\sqrt{2}\,G_F m_W^2\left(1-m_W^2/m_Z^2\right)/\pi$ and the mixing angle
$\cos\theta_W^2=(m_W^2-i\Gamma_W\,m_W)/(m_Z^2-i\Gamma_Z\,m_Z)$, 
using the complex-mass scheme~\cite{Denner:2005fg} throughout.
As input parameters we choose the PDG~\cite{Patrignani:2016xqp} values: 
$G_F = 1.16639\times 10^{-5}$\,GeV$^{-2}$, $m_W=80.385$\,GeV,
$\Gamma_W=2.0854$\,GeV, $m_Z = 91.1876$\,GeV, $\Gamma_Z=2.4952$\,GeV,
$m_H = 125$\,GeV, and $\Gamma_H = 0.00407$\,GeV.
The on-shell masses of bottom and top quarks are  $m_b = 4.92$\,GeV 
and $m_t = 173.2$\,GeV, respectively, and the top width is set to $\Gamma_t=1.44262$\,GeV.
As discussed in the previous section, all predictions are calculated in the 4FS with $n_f=4$ massless quark flavours 
and massive bottom and top quarks, and all contributions with real bottom quarks in the final state are dropped 
to avoid the contamination from top-quark production processes.
Accordingly, we use
 the $n_f=4$ NLO and NNLO sets of the NNPDF3.1 PDFs at NLO(+NLL) and at NNLO(+NNLL/N$^3$LL)~\cite{Ball:2017nwa}, respectively, 
by exploiting the \textsc{Lhapdf} interface~\cite{Buckley:2014ana} with the corresponding values of the strong coupling.
Jets are defined according to the anti-$k_t$ algorithm~\cite{Cacciari:2008gp} with a jet radius of $R=0.4$ and no rapidity requirements.

\renewcommand\arraystretch{1.5}
\begin{table}[t]
\begin{center}
\begin{tabular}{l | c  }
\toprule
\bf lepton cuts
& $\ptlep > 27$\,GeV, \quad $\etalep<2.5$, \quad $\mll>55$\,GeV,  \quad $p_{T,\rm \ell^- \ell^+} >30$\,GeV\\
\bf neutrino cuts
& $\ptmiss > 20$\,GeV\\
\multirow{3}{*}{\bf jet cuts}
 & anti-$k_T$ jets with $R=0.4$;\\
& $N_{\rm jet} = 0$ for $\ptj>35$\,GeV\\
&{\it (our results do not include any cut on $\eta_{\rm jet}$, see text)}\\
\bottomrule
\end{tabular}
\end{center}
\renewcommand{\baselinestretch}{1.0}
\caption{\label{tab:cuts} Fiducial cuts corresponding to the 13 TeV ATLAS analysis~\cite{Aaboud:2019nkz}. See text for details.}
\vspace{0.75cm}
\end{table}
\renewcommand\arraystretch{1}

We employ the following dynamical setting for the central factorisation and renormalisation scales,
\begin{equation}\label{eq:murmuf}
	\mu_F = \mu_R = \mu_0 =\frac{1}{2} \sqrt{\MWW^2+ {\ptWW}^2},
\end{equation}
and set the central resummation scale to
\begin{equation}\label{eq:mufact}
	Q= Q_0 =\frac{1}{4} \MWW,
\end{equation}
where $\MWW$ is the invariant mass of the $W^+W^-$ pair and  $\ptWW$ is its transverse momentum.
Perturbative uncertainties are estimated by performing seven-point scale variations of $\mu_F$ and $\mu_R$ around $\mu_0$ by a factor of two in either direction, keeping $1/2 \leq \mu_F/\mu_R \leq 2$ for $Q=Q_0$, and by varying $Q$ by a factor of two around $Q_0$ in either direction for $\mu_F = \mu_R = \mu_0$.
The total scale uncertainty is evaluated as the envelope of the resulting nine variations.
For the exponent of the modified logarithm in eq~\eqref{eq:modlog} we choose a value of $p=4$ when evaluating the resummed $\ptWW$ spectra
and $p=5$ for the jet-vetoed cross section as well as the $\ptj$ distribution.
We have explicilty checked that the dependence of the results on variations of $p$ is negligible.
Moreover, no non-perturbative corrections are included in our results.

In table~\ref{tab:cuts} we summarise the set of fiducial cuts used in our study.
Those cuts are the same as in the fiducial selection of the 13 TeV analysis of ATLAS in ref.~\cite{Aaboud:2019nkz}, with the only exception that we do not apply any rapidity requirement on the jets since this would 
generate logarithmic corrections of non-global nature, which would spoil the formal accuracy of our resummed results.\footnote{We have checked that the impact of neglecting the rapidity cut on jets is at the permille level for the fiducial cross section.} Furthermore, we introduce two notations: \texttt{fiducial-JV} will be used to refer to the 
full set of fiducial cuts in table~\ref{tab:cuts}, while \texttt{fiducial-noJV} denotes the same fiducial 
setup, but without any restriction on the jet activity.

\subsubsection{Jet-vetoed cross section}\label{sec:pTJveto}
We consider the cumulative cross section in the fiducial phase space as a function of a veto on the transverse momentum of the leading jet ($\ptjv$) 
as generically defined in eq.~\eqref{eq:cumulative}
where we identify the upper bound of the integral with the jet-veto cut $\ptjv$, i.e.\
\begin{align}\label{eq:cumulative-jetveto}
	\sigma (\ptjv) = \int_0^\ptjv \rd \ptj \,\frac{\rd \sigma}{\rd\ptj}\,. 
\end{align}
The corresponding jet-veto efficiency is defined as
\begin{equation}\label{eq:jetvetoeff}
	\varepsilon (\ptjv) = \sigma (\ptjv)/\sigma(\ptjv\rightarrow \infty)\,,
\end{equation}
where $\sigma(\ptjv\rightarrow \infty)$ is the integrated cross section in the \texttt{fiducial-noJV} phase space.

\begin{figure}[t]
  \centering
  \includegraphics[trim={0 -0.2cm 0 
    0},width=0.49\linewidth]{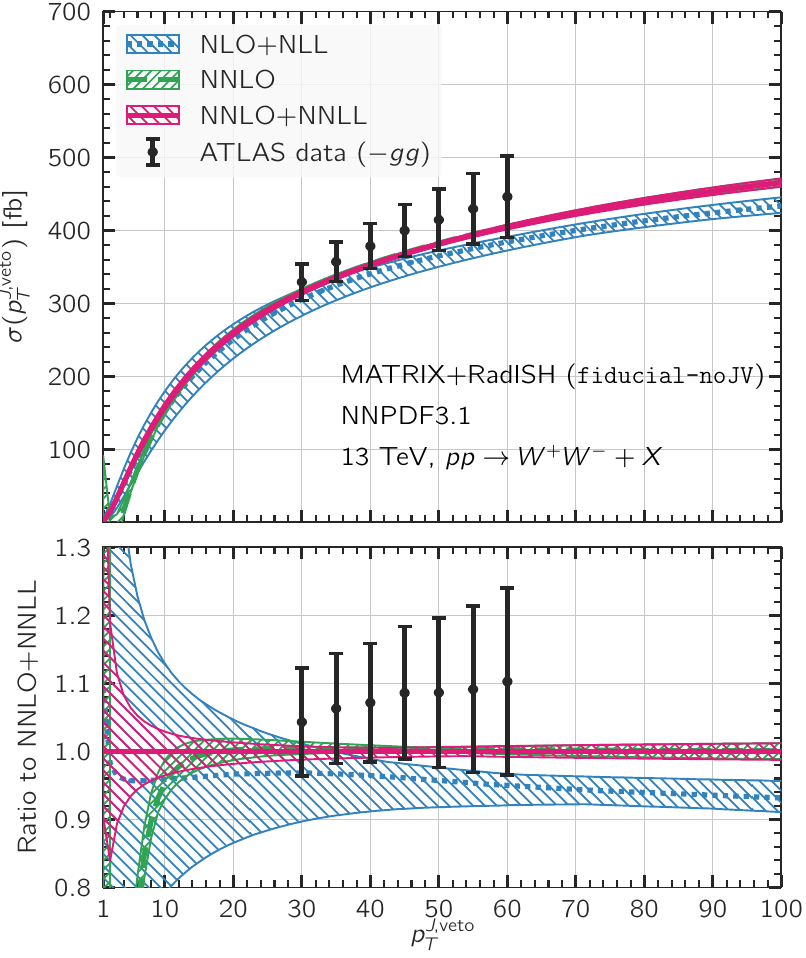} 
  \includegraphics[trim={0 -0.2cm 0
    0},width=0.49\linewidth]{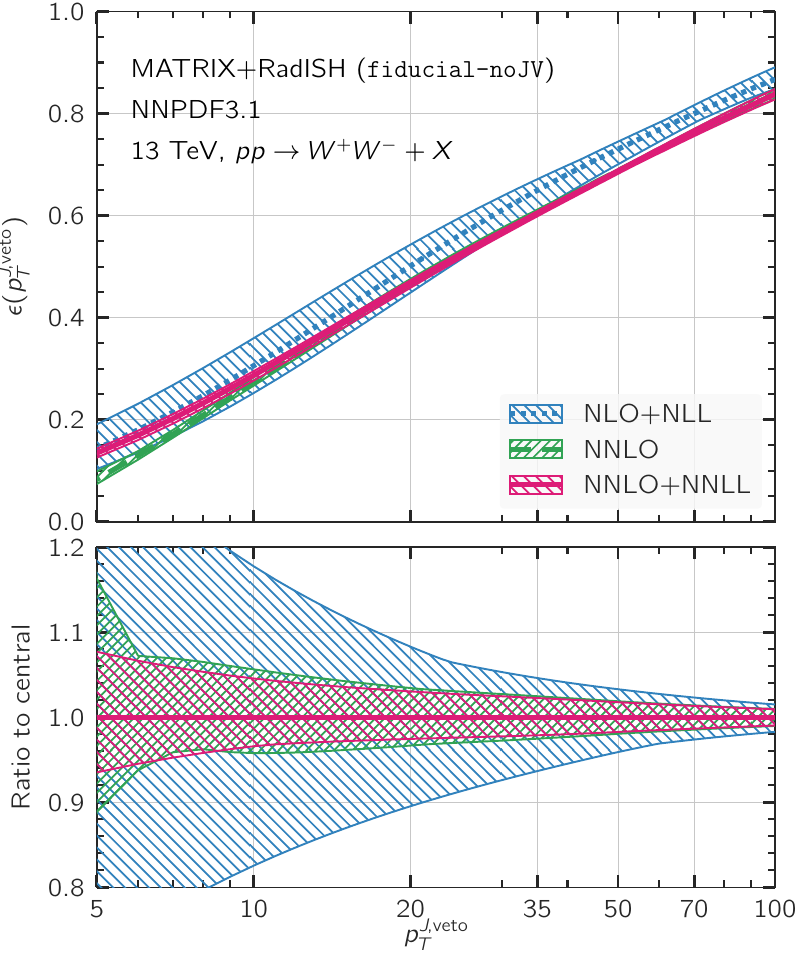} 
  \caption{Fiducial cross sections (left plot) and jet-veto efficiencies (right plot) as a function of the jet-veto cut at NLO+NLL (blue, dotted), NNLO (green, dashed), and NNLO+NNLL (red, solid). In the left plot, we also show the fiducial cross sections measured by ATLAS~\cite{Aaboud:2019nkz}, from which we have subtracted the loop-induced gluon fusion contribution at LO that simply amounts to a constant shift independent of $\ptjv$ (see text for details).}
  \label{fig:ptveto}
\end{figure}

In figure~\ref{fig:ptveto} we show results for the jet-vetoed cross section and the jet-veto efficiency as a function of $\ptjv$.
In the case of the cumulative cross section, we compare our predictions with the cross-section measurements from the ATLAS experiment at $13$\,TeV~\cite{Aaboud:2019nkz}.
Since our resummed predictions do not include the loop-induced gluon fusion contribution, we have subtracted its LO prediction from 
the experimental results to facilitate a meaningful comparison.
We observe that the NNLO result is in remarkable agreement with the NNLO+NNLL prediction down to jet-veto cuts of $\ptjv \gtrsim 15$\,GeV.
Below $\sim 10$\,GeV the fixed-order result becomes unphysical, and the central prediction eventually turns negative.
The fixed-order scale uncertainty band vastly underestimates missing higher-order
corrections since this region is dominated by large logarithms.
The uncertainties of the resummed results increase below $10$\,GeV, and
by comparing NNLO+NNLL and NLO+NLL results it is clear that higher-order corrections in both the fixed-order and the logarithmic expansion
are substantial and mandatory to provide reliable scale uncertainties, reduced to the few-percent level.\footnote{We note that these results resemble closely the comparison between NNLO, NNLOPS and MiNLO in figure~6 of ref.~\cite{Re:2018vac}.}
Due to the relatively large experimental uncertainties, all predictions are in reasonably good agreement with the data.
To better appreciate resummation effects in the comparison with data,
it would be required to push the measurements to much lower jet-veto cuts.

Looking at the jet-veto efficiency in the right plot of figure~\ref{fig:ptveto}, at $35$\,GeV, which is the value of the jet veto used in the fiducial phase space definition, the efficiency is $65$\%, and agreement between the NNLO and the NNLO+NNLL results is at the few permille level. Scale uncertainties for the efficiencies are calculated by considering fully correlated scale variations between the numerator and denominator of eq.~\eqref{eq:jetvetoeff}.
The scale uncertainty at NNLO+NNLL is about $5$\% at $\pt \sim 5$\,GeV, and it decreases towards larger values of the jet veto, being about $2$\% at $\pt \sim 35$\,GeV.
The inclusion of the higher-order corrections reduces significantly the perturbative uncertainties, as illustrated by the NLO+NLL and the NNLO+NNLL bands.

\subsubsection{Differential distributions}\label{sec:diff}

\begin{figure}[t]
  \centering
  \includegraphics[trim={0 -0.2cm 0
    0},width=0.49\linewidth]{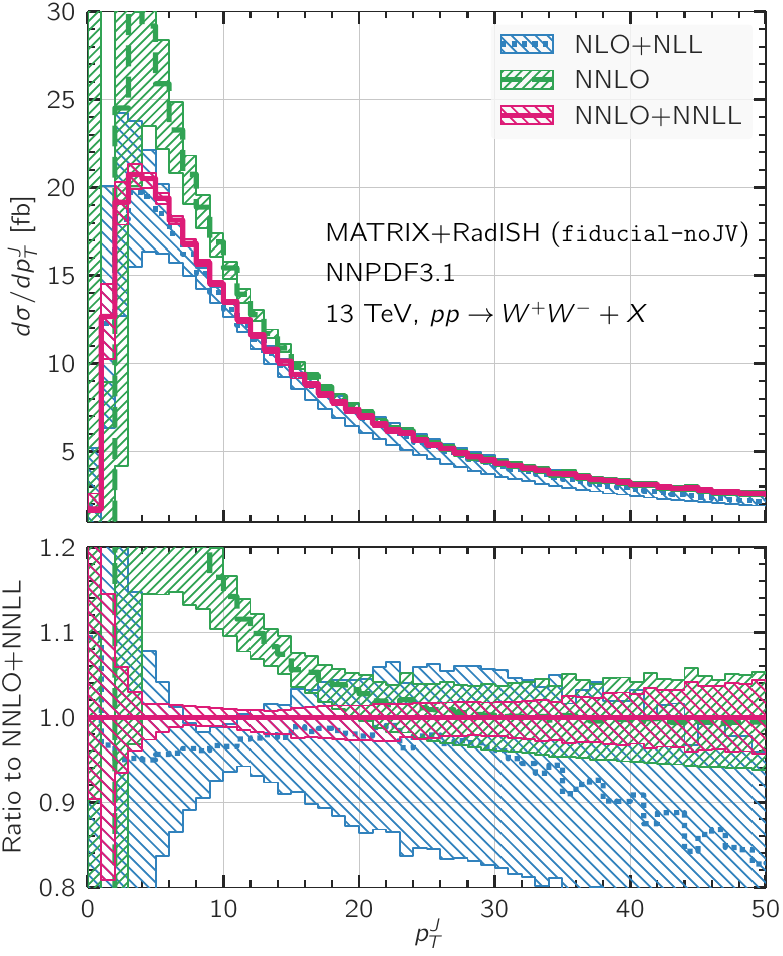} 
  \includegraphics[trim={0 -0.2cm 0
    0},width=0.49\linewidth]{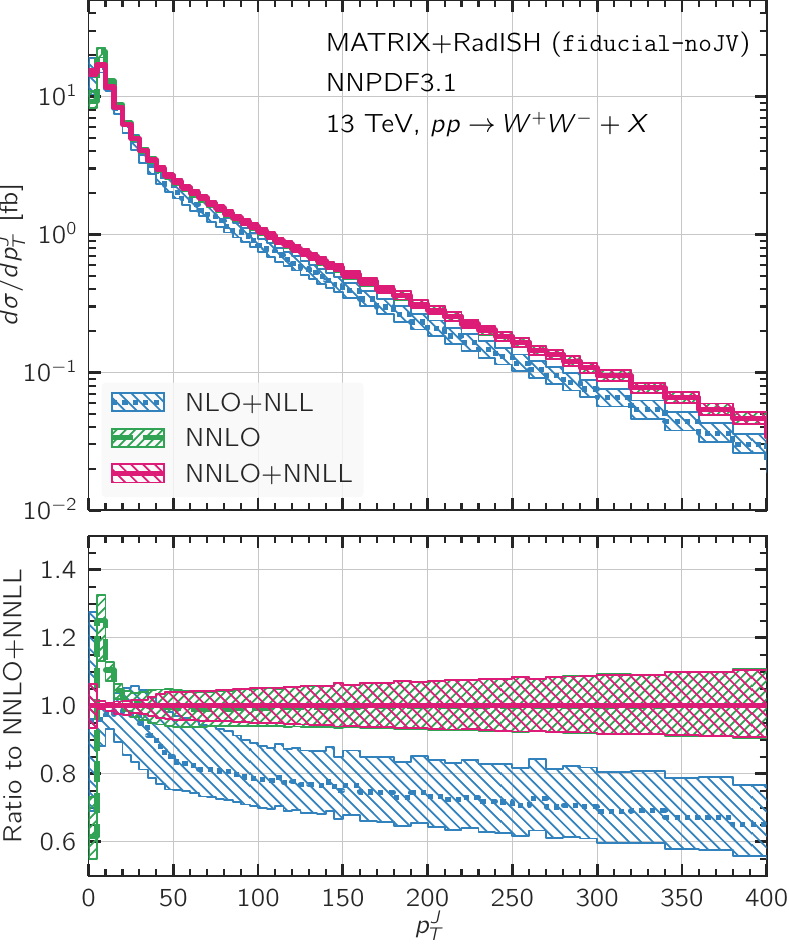} 
  \caption{Transverse-momentum spectrum of the leading jet in the  \texttt{fiducial-noJV} phase space at NLO+NLL (blue, dotted), NNLO (green, dashed), and NNLO+NNLL (magenta, solid) accuracy. Left plot: spectrum up to 50 GeV. Right plot: spectrum up to 200 GeV. The lower frames show the ratios of the 
  predictions to the central value of our best prediction at NNLO+NNLL.}
  \label{fig:ptJ}
\end{figure}

We now move to differential results in the \texttt{fiducial-noJV} phase space.
In figure~\ref{fig:ptJ} we show the transverse-momentum spectrum of the leading jet.
The central value of the NNLO+NNLL prediction lies within the uncertainty band of the NLO+NLL result for $\ptj \lesssim 50$\,GeV.
At larger $\ptj$, where the (N)NLO+(N)NLL result matches the (N)NLO one, the uncertainty bands do not overlap anymore with a visible gap between them in the tail of the distribution.
This indicates that NNLO corrections are particularly important at high values of $\ptj$.
Resummation effects become crucial for $\ptj \lesssim 20$\,GeV, where the NNLO distribution 
is marred by large logarithmic contributions and starts diverging.
The NNLO+NNLL curve is instead well-behaved, and it has perturbative uncertainties of only a few percent down to $\ptj\sim 5$\,GeV.
Below that value, the uncertainty band becomes wider, reaching up to $20\%$ in the first two bins. We note, however, that these regions are likely beyond experimental reach.
In the tail of the NNLO+NNLL prediction the uncertainties gradually increase from about $5\%$ at $\ptj=50$\,GeV to 
about $10\%$ at $\ptj=400$\,GeV.
The Sudakov peak of the resummed $\ptj$ spectrum is around $4$--$5$\,GeV both at NLO+NLL and at NNLO+NNLL.

\begin{figure}[t]
  \centering
  \includegraphics[trim={0 -0.2cm 0
    0},width=0.49\linewidth]{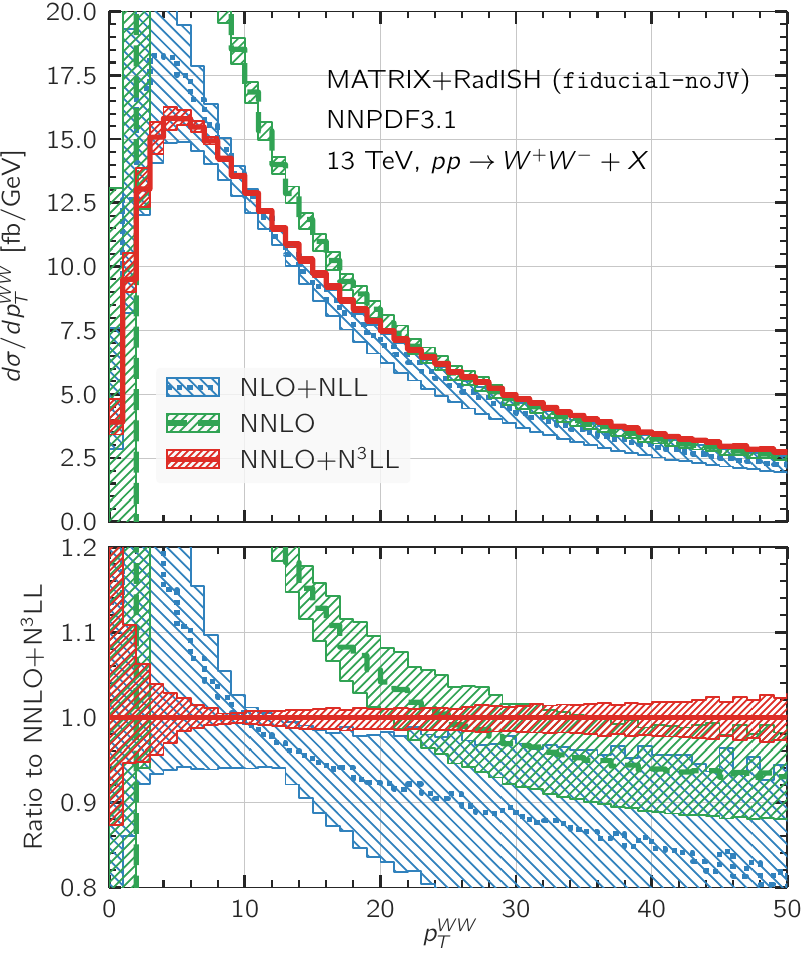} 
  \includegraphics[trim={0 -0.2cm 0
    0},width=0.49\linewidth]{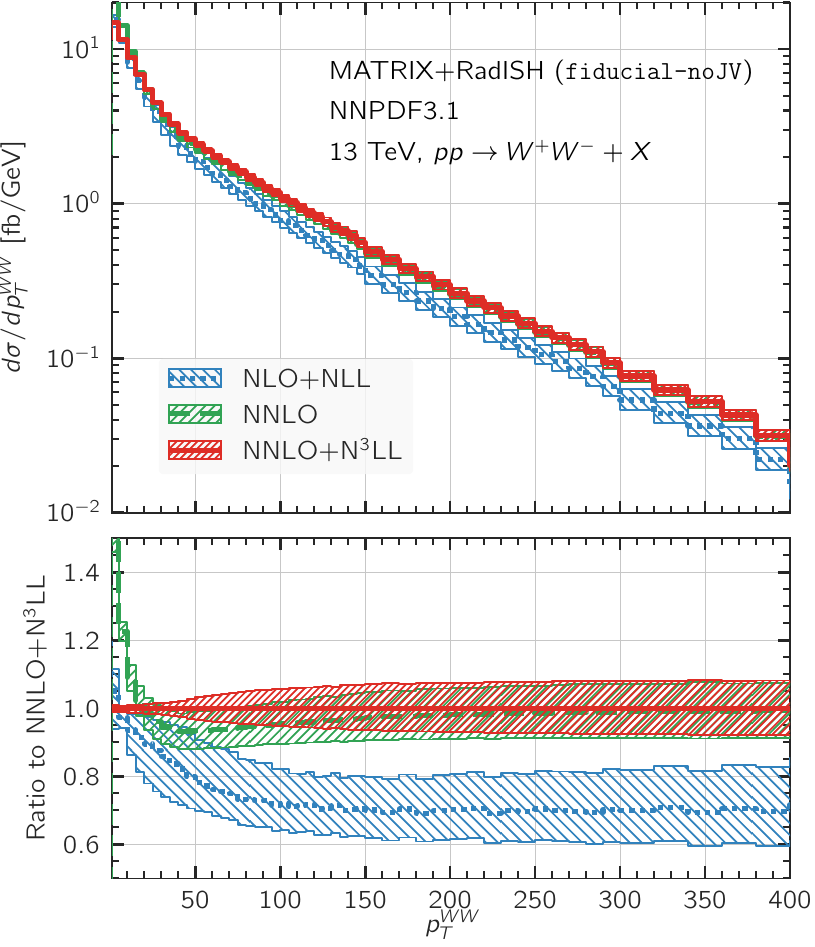} 
  \caption{Transverse-momentum spectrum of the $W^+ W^-$ pair in the  \texttt{fiducial-noJV} phase space at NLO+NLL (blue, dotted), NNLO (green, dashed), and NNLO+N$^3$LL (red, solid) accuracy.  
  Left plot: spectrum up to 50 GeV. Right plot: spectrum up to 200 GeV. The lower frames show the ratios of the 
  predictions to the central value of our best prediction at NNLO+N$^3$LL.}
  \label{fig:ptWW1}
\end{figure}

In figure~\ref{fig:ptWW1} we show the transverse-momentum spectrum of the $W^+W^-$ pair at NNLO as well as matched predictions at NLO+NLL and at NNLO+N$^3$LL.
For $\ptWW \lesssim 20$\,GeV, the NNLO result starts diverging and thus loses predictivity. In that region 
only the matched results are well-behaved since the resummation of large logarithmic contributions becomes indispensable to obtain a physical prediction.
The peak of the NNLO+N$^3$LL spectrum is around $5$--$6$\,GeV, and it is shifted with respect to the NLO+NLL prediction
by about $1$\,GeV, which peaks around $4$--$5$\,GeV.\footnote{We note that these findings are very similar
to the $b$-space results for on-shell $W^+W^-$ production of ref.~\cite{Grazzini:2015wpa}.}
Also at larger values of $\ptWW$, roughly up to $2\,m_W$, the NNLO and NNLO+N$^3$LL curves differ by about $1$\%--$6$\%, 
which indicates that in this intermediate region there are important effects due to the interplay between 
the resummation of large logarithmic contributions and the matching.
The matched result at NNLO+N$^3$LL features very small uncertainties, which are at the few-percent level below $50$\,GeV and gradually increase in the peak region and below ($\ptWW\lesssim 8$\,GeV),
where the logarithmic terms are dominant.
We note that, since the scale variation band is so small, it does not necessarily capture the actual size of  the uncertainties due to missing higher-order terms.
For $20 \lesssim \ptWW \lesssim 50 $\,GeV the scale uncertainties of the matched NNLO+N$^3$LL prediction are only $2$\%--$3$\%, and 
more than a factor of 2 smaller than those of the NNLO result. 
In this region, the uncertainty bands at NNLO and  NNLO+N$^3$LL overlap only marginally.
The importance of higher-order corrections in the fixed-order expansion manifests itself especially
in the tail of the distribution, where the NNLO result is about $30\%$ larger than the NLO+NLL one.

\begin{figure}[t]
  \centering
  \includegraphics[trim={0 -0.2cm 0
    0},width=0.49\linewidth]{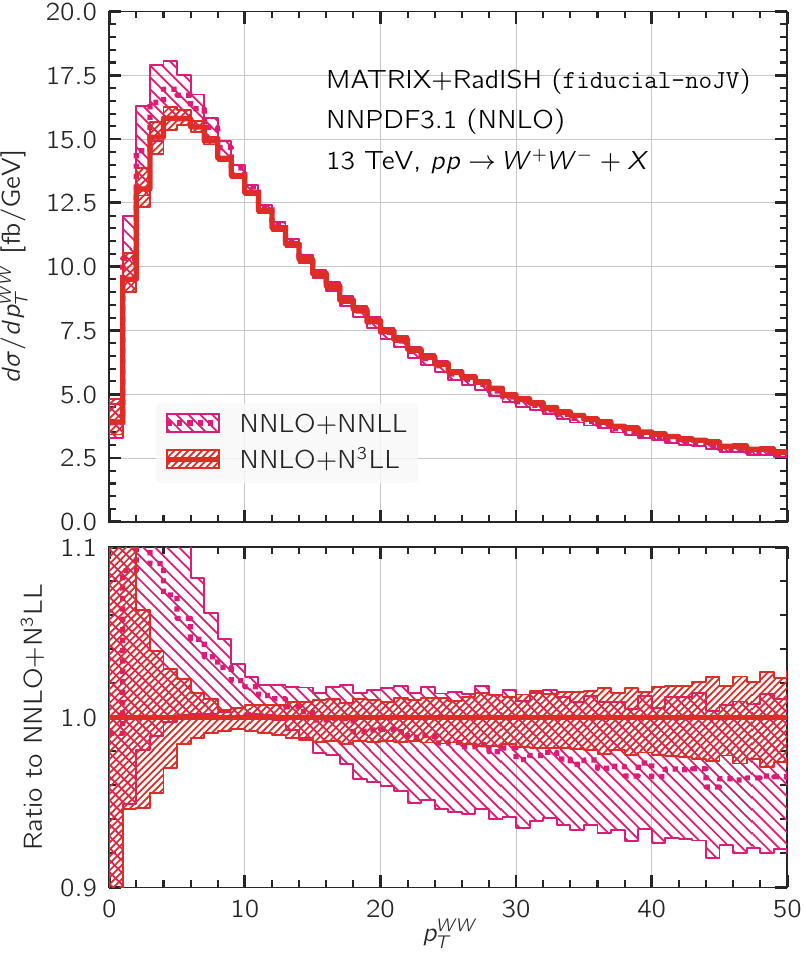} 
  \includegraphics[trim={0 -0.2cm 0
    0},width=0.49\linewidth]{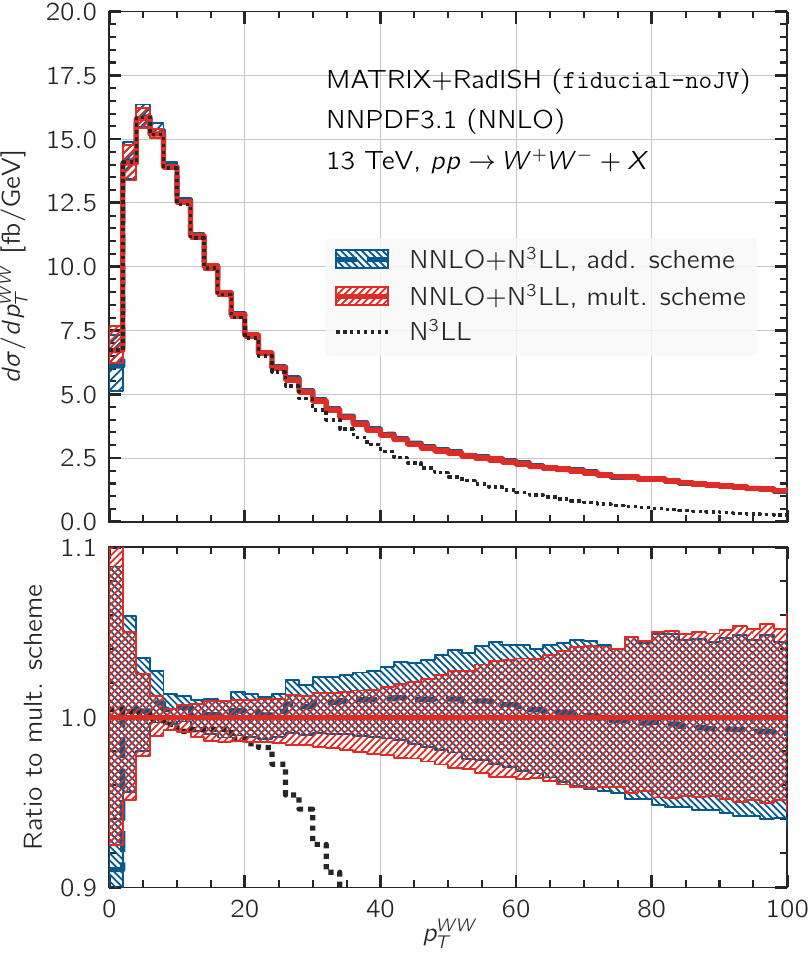} 
  \caption{Left plot: transverse-momentum spectrum of the $W^+ W^-$ pair in the  \texttt{fiducial-noJV} phase space comparing predictions at NNLO+NNLL (magenta, dotted) and NNLO+N$^3$LL (red, solid) accuracy. The lower frame shows the ratios of the 
  predictions to the central value of the NNLO+N$^3$LL result. Right plot: same distribution comparing additive (dark blue, dashed) and multiplicative (red, solid) matching schemes at NNLO+N$^3$LL. The lower frame shows the ratios of the 
  predictions to the central value of the multiplicative result.}
  \label{fig:ptWW2}
\end{figure}

To further study resummation effects in the region $\ptWW \lesssim 50$\,GeV, we compare the result at NNLO+N$^3$LL to the one at NNLO+NNLL in the left plot of figure~\ref{fig:ptWW2}.
The effect of the N$^3$LL corrections on the central prediction is quite sizeable, especially below $10$\,GeV, 
where it reaches almost $10$\%. The uncertainty bands, however, largely 
overlap, with the central prediction at NNLO+N$^3$LL being
fully contained within the NNLO+NNLL band. The inclusion of NNLO+N$^3$LL corrections reduces the scale uncertainties by a bit less than a factor of two.

To illustrate uncertainties due to higher-order effects of the NNLO+N$^3$LL prediction  in the small-$\ptWW$ region, we 
compare the results for two different matching schemes, defined in eqs.~\eqref{eq:additive} and~\eqref{eq:multiplicative1},
in the right plot of figure~\ref{fig:ptWW2}.
At this accuracy, the two predictions contain the same ingredients and are compared on equal footing.
We observe an excellent agreement between the two prescriptions, which indicates that our predictions exhibit a very mild 
dependence on the choice of the matching scheme.
Only at very small transverse momenta ($\ptWW \lesssim 2$\,GeV) we observe minor differences between the multiplicative result and the additive result due to the higher sensitivity of the additive matching to the exact cancellation between the fixed-order 
cross section and the expansion.
The advantage of the multiplicative scheme is confirmed by the fact
that the multiplicative matching 
is in perfect agreement with the pure N$^3$LL result,
as it should be at very small transverse momenta, whereas the additive result is slightly different.
Since the cancellation in the additive matching prescription is numerically challenging in this region,
dedicated runs as those performed in section~\ref{sec:implementation} would be required to achieve a more stable result.

\begin{figure}[t]
  \centering
  \includegraphics[trim={0 -0.15cm 0
    0},width=0.48\linewidth]{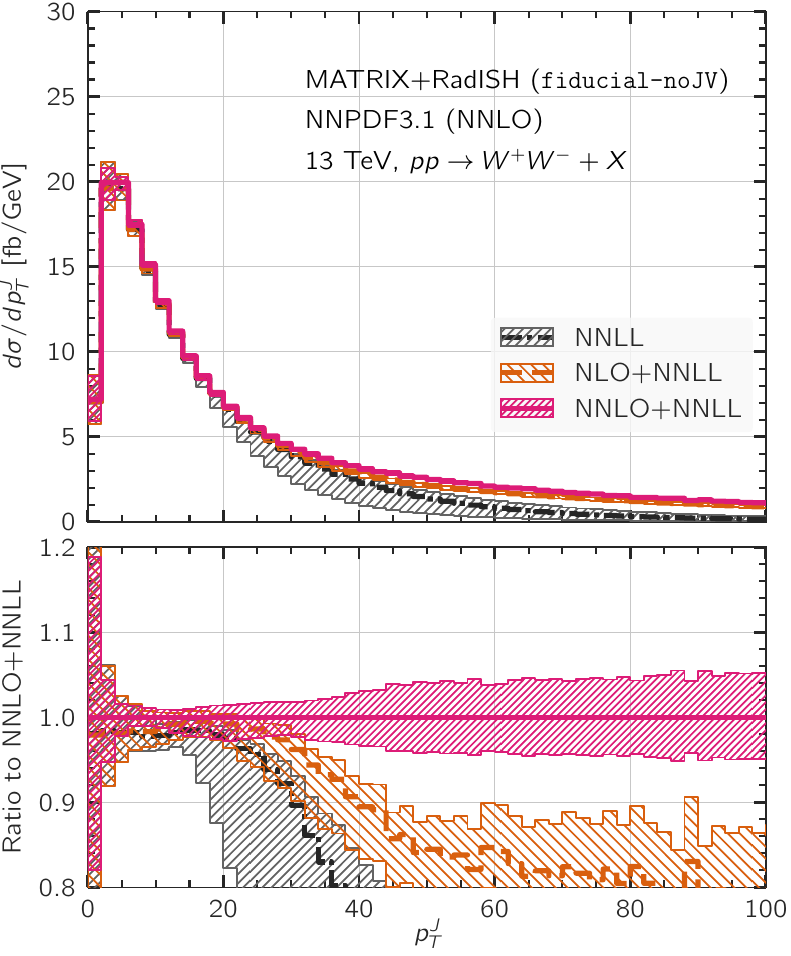} 
  \includegraphics[trim={0 -0.2cm 0
    0},width=0.49\linewidth]{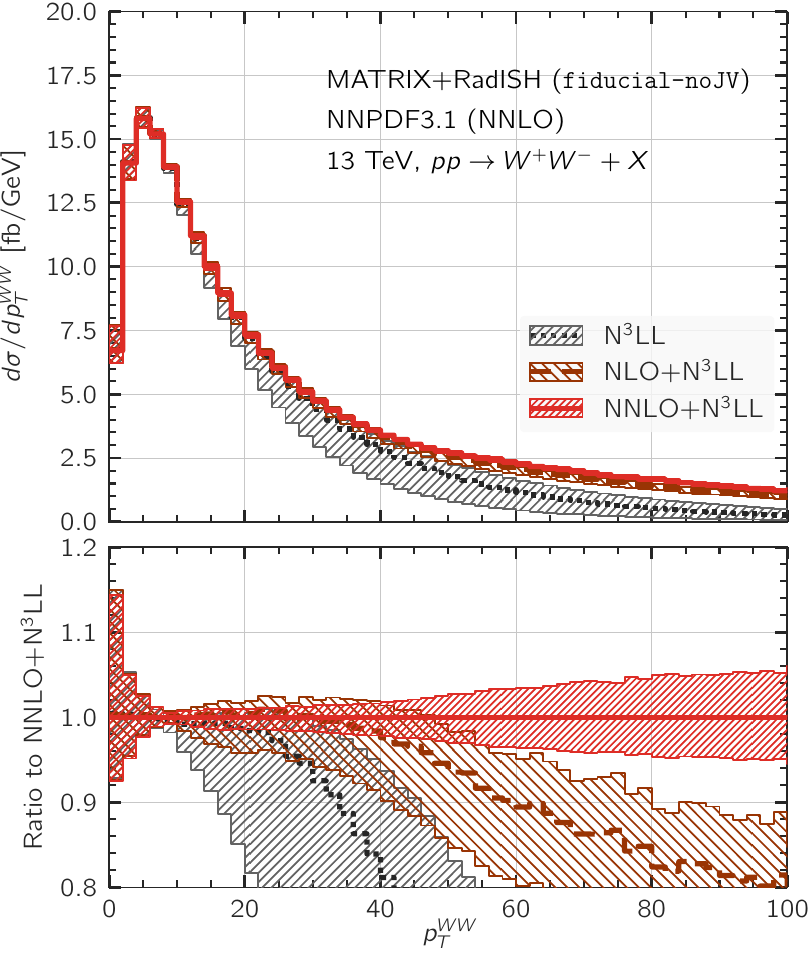} 
  \caption{Left plot: transverse-momentum spectrum of the leading jet in the  \texttt{fiducial-noJV} phase space comparing predictions at NNLO+NNLL (magenta, solid), NLO+NNLL (ochre, dashed) and NNLL (black, dot-dashed) accuracy. The lower frame shows the ratios of the 
  predictions to the central value of the NNLO+NNLL result. Right plot: transverse-momentum spectrum of the $W^+ W^-$ pair in the \texttt{fiducial-noJV} phase space comparing predictions at NNLO+N$^3$LL (red, solid), NLO+N$^3$LL (brown, dashed) and N$^3$LL (black, dotted) accuracy (right plot). 
  The lower frame shows the ratios of the 
  predictions to the central value of the NNLO+N$^3$LL result.}
  \label{fig:ptJ_ptWW}
\end{figure}

So far we have demonstrated the importance of resummation at small transverse momentum and of NNLO corrections
in the tail of the $\ptj$ and $\ptWW$ spectra.
However, the plots in figure~\ref{fig:ptJ} and figure~\ref{fig:ptWW1} are not sufficient to appreciate the impact of the matching to NNLO at small and intermediate values of $\ptj$ and $\ptWW$. 
To this end, in the left (right) plot of figure~\ref{fig:ptJ_ptWW} we 
investigate the impact of the NNLO corrections in the peak and in the 
matching region by comparing NNLL (N$^3$LL) and NLO+NNLL (NLO+N$^3$LL)
predictions to our NNLO+NNLL (NNLO+N$^3$LL) results for $\ptj$ ($\ptWW$). Below $\sim 20$\,GeV matching effects play a minor role, while beyond 
this value the non-singular corrections become large and the purely resummed result unreliable.
Note that for $\ptj\lesssim 20$\,GeV there is a moderate 
increase of $\sim 2\%$ originating from the inclusion of NNLO corrections in 
the matching. This difference can be traced back to the the NNLO constant terms
in $\ptj$, which are absent in the NNLL result and are included through the 
multiplicative matching at small $\ptj$, as discussed in section~\ref{sec:implementation}. That behaviour
is not observed for $\ptWW$, since the N$^3$LL result already includes the 
NNLO constant terms. Looking at the matching region, the 
inclusion of NNLO corrections becomes important for $\ptj \gtrsim 30$\,GeV
and $\ptWW \gtrsim 40$\,GeV. Already at $\ptj\sim 35$\,GeV and $\ptWW\sim 50$\,GeV the uncertainty bands of the predictions matched to NLO and to NNLO 
do not overlap anymore. Not least, the matching to NNLO has a substantial impact on the size of the uncertainty bands, which above $\ptj\sim 25$\,GeV and $\ptWW\sim 10$\,GeV are reduced by roughly a factor of two. In conclusion, 
NNLO accuracy plays an important role not only for the accurate
description of the high-$\pt{}$ tail, but also in the matching region.

\begin{figure}[t]
  \centering
  \includegraphics[trim={0 -0.2cm 0
    0},width=0.49\linewidth]{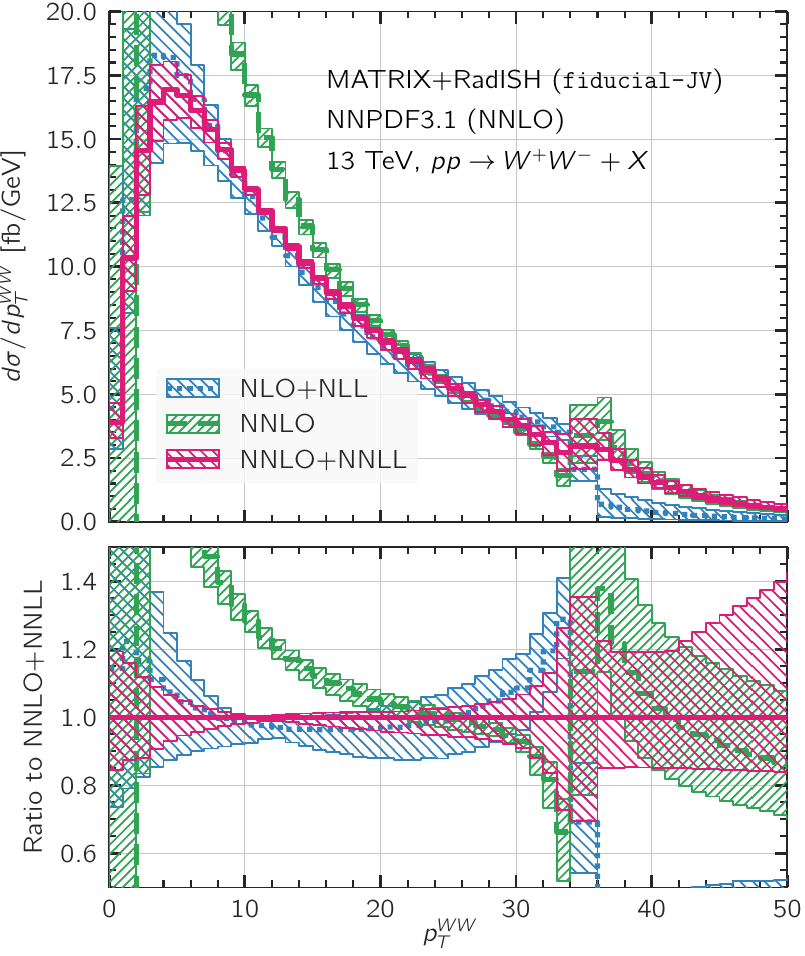} 
  \includegraphics[trim={0 -0.2cm 0
    0},width=0.49\linewidth]{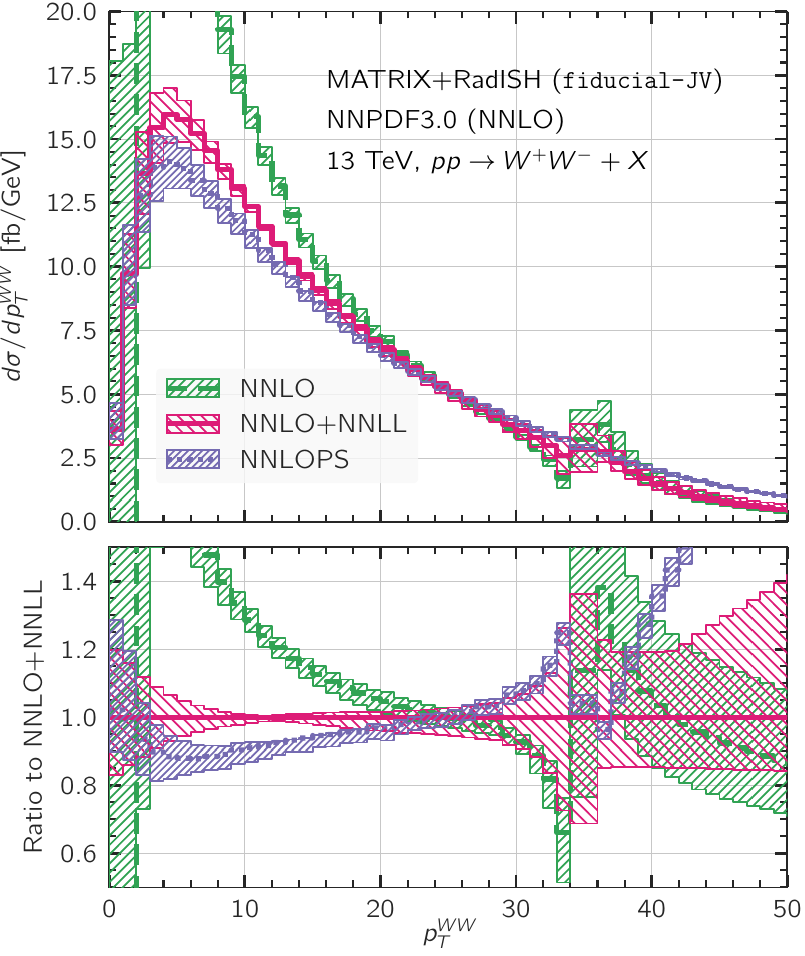} 
  \caption{Left plot: transverse-momentum spectrum of the $W^+ W^-$ pair with a jet-veto requirement in the \texttt{fiducial-JV} phase space at NLO+NLL (blue, dotted), NNLO (green, dashed), and NNLO+NNLL (red, solid) accuracy. Right plot: same distribution comparing NNLO (green, dashed) and NNLO+NNLL (red, solid) predictions to the NNLOPS result (blue, dotted) of ref.~\cite{Re:2018vac}. To facilitate the comparison, we reanalysed the NNLOPS events and we recalculated our predictions with the NNPDF3.0 NNLO PDF set~\cite{Ball:2014uwa}. The lower frame shows the ratios of the 
  predictions to the central value of the NNLO+N$^3$LL result.}
  \label{fig:JV_ptWW}
\end{figure}

We continue our studies by considering the transverse-momentum spectrum of the $W^+W^-$ system
in presence of a jet veto. To this end, we perform the joint resummation of large logarithmic terms
in both $\ptWW$ and $\ptj$. 
In the left plot of figure~\ref{fig:JV_ptWW}, we compare resummed predictions for the $\ptWW$ spectrum
in the \texttt{fiducial-JV} phase space at NLO+NLL and NNLO+NNLL accuracies to the NNLO result. By exploiting the multiplicative matching scheme at the double-differential level, defined in eq.~\eqref{eq:multiplicative2}, we include the constant terms in the resummed prediction, and the integral of the NLO+NLL (NNLO+NNLL) $\ptWW$ spectrum 
yields the NLO+NLL (NNLO+NNLL) jet-vetoed cross section.
The NNLO curve develops a perturbative instability~\cite{Catani:1997xc} right at $\ptWW=35$\,GeV, which 
corresponds to the value of the jet-veto cut. This instability is caused by an incomplete cancellation between soft contributions 
in the real and virtual amplitudes, and it leads to an integrable logarithmic divergence at the threshold. Since $\ptWW=\ptj$ holds at LO, and consequently $\ptWW<35$\,GeV, the $\ptWW$ region above the jet-veto cut is filled only 
by higher-order corrections, and the effective perturbative accuracy is reduced by one order. This is indicated 
by the widening of the uncertainty band of the NNLO prediction for $\ptWW>35$\,GeV, which effectively becomes 
NLO accurate in that region.

The sensitivity to the perturbative instability at the threshold is largely cured by the NNLO+NNLL
result, resumming a large part of the relevant logarithms.
However, a slight sensitivity remains due to the fact that our approach resums Sudakov logarithms in
the limit where $\ptWW$ and $\ptj$ are much smaller
than the hard scale, while additional logarithmic terms
contribute when hard jets are present.
Nevertheless, the large differences between the 
NNLO+NNLL and NNLO results indicate the importance of resummation at the threshold.
We observe even larger resummation effects for transverse momenta below $20$\,GeV, where the 
NNLO result becomes unphysical and resummation is mandatory to retain a reliable prediction.
The resummed spectra at NLO+NLL and NNLO+NNLL are consistent with each other at small $\ptWW$,
with fully overlapping uncertainty bands up to \mbox{$\ptWW\sim 30$\,GeV}.

The theoretical uncertainty of the NNLO+NNLL prediction is at the few-percent level in that region and roughly a factor of two smaller than the NLO+NLL uncertainty band. It reaches 
$\mathcal O(10\%)$ uncertainties only for $\ptWW\lesssim 6$\,GeV, where the logarithmic corrections become larger.
When approaching the perturbative instability, the differences between NLO+NLL and NNLO+NNLL progressively
increase, with barely overlapping uncertainties below the threshold. 
At the threshold theory uncertainties of the NNLO+NNLL result reach $30\%$.
The large differences between the resummed results and the widening of the uncertainty bands further
indicate that additional logarithmic terms contribute at the jet-veto threshold, which are not resummed.
Beyond threshold the NLO+NLL result becomes unreliable, being effectively only LO accurate, while the NNLO+NNLL 
prediction is effectively NLO accurate with an enlarged uncertainty band at the $20$\%--$30$\% level.

In conclusion, our results indicate that the inclusion of higher-order corrections both in the 
fixed-order and the logarithmic expansion is particularly relevant for an accurate 
description of the $\ptWW{}$ distribution in presence of a jet veto.
To further study resummation effects for this observable, we compare the results at 
NNLO+NNLL with the NNLOPS results of ref.~\cite{Re:2018vac} in figure~\ref{fig:JV_ptWW}.
By and large, the two results are in reasonable agreement. At very small $\ptWW$ their uncertainty bands
overlap. There is a gap between them for $5\,\text{GeV}\lesssim\ptWW\lesssim 15$\,GeV,
because the NNLO+NNLL band almost vanishes at $\ptWW\sim 10$\,GeV, 
and thus underestimates the actual uncertainties in that region, and 
also the NNLOPS band,
which misses uncertainties related to the shower-starting scale, 
is somewhat underestimated. For $\ptWW\lesssim 25$\,GeV, the NNLO+NNLL result 
provides the more accurate and therefore more reliable prediction.
When approaching the threshold at $35$\,GeV, the two results agree within uncertainties due to 
the rather large NNLO+NNLL band. The NNLOPS result is smooth in that region 
and develops no perturbative instability.
At higher $\ptWW$ values, the NNLOPS prediction
becomes significantly larger than the NNLO+NNLL result since the shower generates additional QCD radiation, as discussed in ref.~\cite{Re:2018vac}.
The comparison shows that in the low-$\ptWW$ region high-accuracy resummation is required for
a precise prediction of the spectrum. For values of $\ptWW$ at and above the threshold, the NNLOPS result gives a more reliable description of the spectrum in the fiducial region, as it includes effects that are not included in other approaches, albeit with a limited formal accuracy.


\section{Summary}\label{sec:conclusion}

In this paper we have introduced a general framework for the computation of accurate cross-section predictions including multi-differential resummation. 
By developing an interface between the codes \textsc{Matrix} and \textsc{RadISH} we have combined fully differential NNLO QCD predictions
for \mbox{$2 \rightarrow 1$} and \mbox{$2 \rightarrow 2$} colour-singlet production processes with high-accuracy all-order resummation.
In particular, \textsc{Matrix+RadISH} evaluates the transverse-momentum spectrum of colour singlets and the $\phs$ distribution for the Drell--Yan process up to NNLO+N$^3$LL accuracy,
the transverse-momentum spectrum of the leading jet and jet-veto resummation up to NNLO+NNLL accuracy, as well as the joint resummation of the colour-singlet and the leading-jet transverse 
momentum up to NNLO+NNLL.
Thereby we provide a powerful and versatile parton-level Monte Carlo generator
allowing for an accurate description of transverse observables in colour-singlet 
production processes with arbitrary cuts on the Born kinematics. 
This framework can be extended to describe any observable 
differential in the Born kinematics including resummation effects by exploiting a suitable scheme for the kinematic recoil within the resummation. Moreover, the framework
facilitates the combination of future developments within \textsc{Matrix} and \textsc{RadISH}, such as advancements 
towards processes beyond colour-singlet production.\sloppy

As a first phenomenological application we have studied the production of $W^+ W^-$ pairs at the LHC. More precisely, 
the full leptonic final state of two charged different-flavour leptons and two neutrinos has been considered, 
including spin correlations, interferences and off-shell effects.
We have presented resummed results at $13$\,TeV for several kinematic distributions in presence of fiducial cuts on the leptonic final states and the associated QCD radiation.
The inclusion of higher-order corrections in both the logarithmic and the fixed-order expansion turns out to be 
essential to achieve a precise description of differential observables at the few-percent level, as demanded by the experimental analyses.
The accurate modelling of the cross section in presence of a veto against hard QCD radiation is particularly important in that respect
since a jet veto is commonly employed in $W^+ W^-$ measurements to suppress top-quark backgrounds. 
Our NNLO+NNLL result yields residual uncertainties at the few-percent level, we have shown that NNLO predictions are reliable down to jet-veto cuts of roughly $15$\,GeV, and we find agreement within one standard deviation with the cross section measured by ATLAS as a function of the jet-veto cut between $30$\,GeV and $60$\,GeV.

For the differential spectra in the transverse momentum of the leading jet and the $W^+W^-$ pair, which we have evaluated
up to NNLO+NNLL and NNLO+N$^3$LL accuracy, respectively, we obtain scale uncertainties that are generally below 5\%.
Furthermore, our results highlight the importance of high-accuracy resummation in the region of the 
Sudakov peak, and of the $\mathcal{O}(\as^2)$ corrections in the tail of the distribution. Indeed, scale variations at $\mathcal{O}(\as)$ significantly 
underestimate the actual size of $\mathcal{O}(\as^2)$ corrections at large transverse momenta.
At small transverse momenta, N$^3$LL corrections to the $W^+W^-$ transverse-momentum 
distribution are still quite sizable, with differences of about 5\% --10\% when comparing NNLO+N$^3$LL  to NNLO+NNLL.

The matching of the resummed cross section, valid in the soft-collinear region, and the fixed-order predictions, valid in 
the hard region, can be achieved in different ways. We have shown that a multiplicative matching procedure,
which at NNLL resums additional logarithmic contributions originating from the NNLO terms that are constant in the resummation variable, has the 
further advantage of being numerically more stable at small transverse momenta than the additive matching procedure. 
Furthermore, we found that at NNLO+N$^3$LL there is
essentially no dependence of the $W^+W^-$ transverse-momentum spectrum on the choice of the matching scheme.

Finally, we have studied the transverse-momentum distribution of $W^+W^-$ pairs in presence of a $35$\,GeV jet-veto requirement
by simultaneously resumming both classes of logarithmic terms. 
At fixed order no realistic description of this observable can be obtained.
Resummed results at NLO+NLL and NNLO+NNLL indicate good perturbative convergence at small transverse momenta (below $\sim 30$\,GeV), and the 
inclusion of the NNLO+NNLL corrections decreases scale uncertainties 
by roughly a factor of two in that regime. The most delicate region is 
the threshold where the $W^+W^-$ transverse momentum is close to the value of the jet-veto cut since this region is subject to a
perturbative instability at fixed order in $\as$.
The joint resummation of jet-veto logarithms 
and logarithmic terms in the $W^+W^-$ transverse momentum
improves the stability of the spectrum in the threshold region
by resumming part of the relevant logarithms.
We have further compared our NNLO+NNLL results 
to NNLOPS predictions, which corroborates the importance of
NNLL resummation for an accurate modeling at small transverse momenta. In the threshold region the NNLOPS result is smooth, and we found it to be in reasonable agreement with NNLO+NNLL predictions, well within the respective scale uncertainties.

In all presented results we refrained from including the loop-induced gluon 
fusion contribution, which is formally part of the NNLO corrections,
because it is effectively only LO accurate and Born-like.
Although at fixed order it contributes trivially to all observables we have considered,
i.e.\ as a constant shift in the cross sections, at the resummed level it is of the same size as N$^3$LL corrections to the quark channel. Since their resummation
can be considered completely independently, we leave 
a proper treatment of the loop-induced gluon fusion contribution, and specifically
the combination of its NLO corrections with NNLL resummation, for future work.

We reckon that the predictions presented in this paper for the specific 
case of $W^+W^-$ production as well as the 
\textsc{Matrix+RadISH} framework in general will be a very useful 
addition to current fixed-order and parton-shower predictions and tools.\footnote{The \textsc{Matrix+RadISH} code is publicly available on \url{https://matrix.hepforge.org/}.}
With this framework we hope to
advance the sensitivity to transverse observables in colour-singlet processes
for precision measurements and new-physics searches 
at the LHC and future $pp$ colliders.

\acknowledgments We thank Massimiliano Grazzini, Pier Monni and Paolo Torrielli for fruitful discussions and useful comments on the manuscript. 
We are grateful to CERN, Max-Planck-Institut f\"ur Physik, and Universit\`a degli Studi di Milano-Bicocca, where part of this project was carried out, for hospitality.
The work of SK and LR is supported by the ERC Starting Grant 714788 REINVENT.
SK and LR acknowledge the CINECA award under the ISCRA initiative for the availability of high-performance computing resources needed for this work.


\addtocontents{toc}{\protect\setcounter{tocdepth}{1}}
\appendix

\section{How to use \textsc{Matrix+RadISH}}\label{app:example}

The aim of this appendix is to provide some guidance on the usage of the \textsc{Matrix+RadISH} interface.
Beside some additional parameters in the input files, mostly related to
\textsc{RadISH}, and some necessary modifications to the structure of the output to accommodate the additional histograms, a \textsc{Matrix+RadISH} run is very similar to a fixed-order run in \textsc{Matrix}.
Therefore, we shall describe only the main steps needed to produce 
resummed predictions and provide only basic information on 
\textsc{Matrix} here, while focussing on 
the aspects that are different in \textsc{Matrix+RadISH}.
We refer the reader to ref.~\cite{Grazzini:2017mhc} for a comprehensive 
overview of all the settings available in \textsc{Matrix} and a complete description of the structure of the code.
If the reader has never run a fixed-order computation with~\textsc{Matrix}, we encourage them 
to consult the \textsc{Matrix} manual in ref.~\cite{Grazzini:2017mhc}, and in particular
section~3 therein, before reading this appendix.

\subsection{Compilation and setup of a process}\label{compiling-the-code}

We assume that the  \texttt{MATRIX+RadISH\_v1.0.0.tar.gz} package has been downloaded and extracted, and that a working installation of \textsc{Lhapdf}~\cite{Buckley:2014ana} is present on the machine such that the shell command \texttt{lhapdf-config} is recognised when it is run in a terminal.
We also require a working \textsc{Python\,2.7} installation.\footnote{Currently the scripts necessary to run the \textsc{Matrix} shell are not yet compliant with \textsc{Python\,3} standards.}
If these requirement have been met, the command
\begin{Shaded}
\begin{Highlighting}[]
  \ExtensionTok{\$  ./matrix}\NormalTok{ --radish}
\end{Highlighting}
\end{Shaded}
\noindent executed inside the  \texttt{MATRIX+RadISH\_v1.0.0} folder will launch the \textsc{Matrix} shell with the \textsc{RadISH} interface enabled.\footnote{For a comprehensive list of the additional options of the \texttt{./matrix} command see section 3.2 of ref.~\cite{Grazzini:2017mhc}.}
An interactive \textsc{Python} session for the compilation will appear, with instructions printed on the screen.
This is followed by the usual steps in the \textsc{Matrix} shell to select a process via its \texttt{\$\{process\_id\}},\footnote{The resummation is available for all colour-singlet processes contained in the public release of \textsc{Matrix}.}
 e.g.\ for Drell--Yan production
\begin{Shaded}
\begin{Highlighting}[]
  \ExtensionTok{|===>> ppz01}
\end{Highlighting}
\end{Shaded}
\noindent and to agree with the terms to use \textsc{Matrix+RadISH} by typing (a few times)
\begin{Shaded}
\begin{Highlighting}[]
  \ExtensionTok{|===>> y}
\end{Highlighting}
\end{Shaded}
\noindent This requires you to acknowledge the work of various groups that 
went into the computation performed by \textsc{Matrix+RadISH} for the present process by citing the references collected in the file 
 \texttt{CITATION.bib}, which is provided with the results in every run.
The compilation shell will then execute the following steps:
\begin{itemize}
\item linking to \textsc{Lhapdf}~\cite{Buckley:2014ana};
\item download and installation of \textsc{OpenLoops} \cite{Cascioli:2011va,Buccioni:2017yxi,Buccioni:2019sur} (skipped if already installed);
\item installation of {\sc CLN} \cite{Haible:1998xxx} (skipped if already installed);
\item installation of {\sc GiNaC} \cite{Bauer:2000cp} (skipped if already installed);
\item installation of {\sc Hoppet} \cite{Salam:2008qg} (skipped if already installed);
\item installation of {\sc Chaplin} \cite{Buehler:2011ev} (skipped if already installed);
\item installation of {\sc RadISH} \cite{Monni:2016ktx,Bizon:2017rah,Monni:2019yyr} (skipped if already installed);
\item compilation of \textsc{Matrix} process (asked for recompilation if executable exists);
\item download of the relevant tree-level and one-loop amplitudes through \textsc{OpenLoops} (skipped if they already exist);
\item setting up of the process folder under the path \texttt{run/\$\{process\_id\}\_MATRIX}\,.
\end{itemize}
Once completed, the shell will exit and the process is ready to be run from the created process folder. As instructed on screen, 
enter that folder (\texttt{cd run/\$\{process\_id\}\_MATRIX}) and start a run by following the instructions in the next section.

\subsection{Running a process in the interactive shell}\label{running-the-code}

Once the process has been compiled, we launch the running shell with the usual command
\begin{Shaded}
\begin{Highlighting}[]
\ExtensionTok{\$ ./bin/run_process}
\end{Highlighting}
\end{Shaded}
\noindent inside the folder (\texttt{cd run/\$\{process\_id\}\_MATRIX}). With this interactive steering interface 
all run-related settings, inputs, and options will be handled just as in a standard \textsc{Matrix} run.\footnote{The script accepts several optional arguments, which are discussed in detail in section~3.5 of ref.~\cite{Grazzini:2017mhc}.}
After typing a name for the run, e.g.
\begin{Shaded}
\begin{Highlighting}[]
  \ExtensionTok{|===>> run_name_resummation}
\end{Highlighting}
\end{Shaded}
\noindent a list of several commands is printed on the screen. Before starting the run, the only differences with respect 
to an ordinary \textsc{Matrix} run are the inputs of the files \texttt{parameter.dat} and \texttt{distribution.dat} of the 
current run inside the folder \texttt{input}. We open them by typing
\begin{Shaded}
\begin{Highlighting}[]
  \ExtensionTok{|===>> parameter}
  \ExtensionTok{|===>> distribution}
\end{Highlighting}
\end{Shaded}
\noindent 
and we modify the settings to turn on the resummation as discussed in appendix~\ref{sec:input-settings}. 
Once the input cards have been adjusted to obtain the desired results, the run is started through the usual command:
\begin{Shaded}
\begin{Highlighting}[]
  \ExtensionTok{|===>> run }
\end{Highlighting}
\end{Shaded}
\noindent From now on, no human intervention is needed. 
Once the run is finished, the results from the run are collected in the respective folder \texttt{result/\$\{run\_name\_resummation\}}, which is printed on screen at the end of the run.

The only difference, compared to an ordinary fixed-order \textsc{Matrix} run, after a run with resummation has been completed, is that 
there will be additional results inside the folder \texttt{result/\$\{run\_name\_resummation\}}. In particular, there will be the following information:
\begin{itemize}
	\item The same results as in a fixed-order run are saved, and the general structure is identical.
	\item If resummation for the respective orders has been turned on in the file \texttt{parameter.dat}, 
as discussed in appendix~\ref{sec:parameters-settings}, inside the folders \texttt{NLO-run} and \texttt{NNLO-run} a folder \texttt{MATRIX+RadISH} is present.
	\item  The folder \texttt{MATRIX+RadISH} contains the resummed differential distributions (ending with \texttt{NLL.dat}, \texttt{NNLL.dat} or \texttt{N3LL.dat}), the corresponding fixed-order differential distributions (ending with \texttt{NLO\_QCD.dat} or \texttt{NNLO\_QCD.dat}), the matched distributions at differential level (ending with \texttt{NLO+NLL.dat}, \texttt{NNLO+NNLL.dat} or \texttt{NNLO+N3LL.dat}), which
	correspond to the best and final prediction, and, for 
completeness, the expansion of the resummed cross section at the relevant order (ending with \texttt{exp\_NLO\_QCD.dat} or \texttt{exp\_NNLO\_QCD.dat}).
	\item There is a further subfolder \texttt{cumulant} inside \texttt{MATRIX+RadISH} that contains the respective
	cumulative cross sections obtained from the differential distributions.
\end{itemize}
 
\subsection{Additional settings in the configuration file}\label{sec:configuration-settings}

\begin{table}
\begin{center}
\resizebox{\columnwidth}{!}{%
\begin{tabular*}{\textwidth}{llplpl}
\toprule
{\bf variable}  & \multicolumn{2}{p{10.2cm}}{\raggedright \bf description}    \\
\midrule
\texttt{path\_to\_radish} & \multicolumn{2}{p{10.5cm}}{\raggedright Path to the \textsc{RadISH} installation; not required in most cases.}\\
\texttt{path\_to\_hoppet} & \multicolumn{2}{p{10.5cm}}{\raggedright Path to \texttt{hoppet-config}; not required in most cases.}\\
\texttt{path\_to\_chaplin} & \multicolumn{2}{p{10.5cm}}{\raggedright Path to the \textsc{Chaplin} installation; not required in most cases.}\\
\bottomrule
\end{tabular*}
}
\end{center}
\renewcommand{\baselinestretch}{1.0}
\caption{Additional parameters available the file \texttt{MATRIX\_configuration}.} 
 \label{tab:config}
\end{table}
\renewcommand\arraystretch{1.1}

Before turning to physics-related and technical settings relevant for a \textsc{Matrix+RadISH} run in the next section, the additional parameters in the file \texttt{MATRIX\_configuration} are discussed. The main file \texttt{MATRIX\_configuration} can be found in the folder \texttt{config} 
inside the \textsc{Matrix+RadISH} main folder. This file is linked during each setup of a process (see appendix~\ref{compiling-the-code}) into the folder
\texttt{input} of the respective process folder.
The additional options for a \textsc{Matrix+RadISH} run in the file \texttt{MATRIX\_configuration} are listed in table~\ref{tab:config}.

\subsection{Additional settings in {\tt parameter.dat} and {\tt distribution.dat}}\label{sec:input-settings}

The various parameters in the input files are described in section 4.1 of ref.~\cite{Grazzini:2017mhc}.
Here we will discuss only the additional settings in the files \texttt{parameter.dat} and \texttt{distribution.dat} 
that are relevant when performing a run in \textsc{Matrix+RadISH}.

\subsubsection{Settings in {\tt parameter.dat}}\label{sec:parameters-settings}

All main parameters, related to the run itself or the behaviour of the code, 
are specified in the file \texttt{parameter.dat}. Most of them should be completely 
self-explanatory, and we will focus our discussion on the essential ones
with a direct connection to the resummation. The settings are organized into certain groups. 
Here, we will limit the discussion to the groups which contain additional information when the \textsc{Matrix+RadISH} interface is enabled, for the sample case of $Z\gamma$ production (where applicable) as in ref.~\cite{Grazzini:2017mhc}.

\newpage
\paragraph{Scale settings}\label{sec:scale}
\lstset{basicstyle=\tiny, frame=single}
{\tt 
\begin{lstlisting}
scale_ren       =  91.1876     #  renormalization (muR) scale
scale_fact      =  91.1876     #  factorization (muF) scale
dynamic_scale   =  6           #  dynamic ren./fac. scale (not working with resummation)
                               #  0: fixed scale above
                               #  1: invariant mass (Q) of system (of the colourless final states)
                               #  2: transverse mass (mT^2=Q^2+pT^2) of system (of the colourless final states)
                               #  3: transverse mass of photon (note: mT_photon=pT_photon)
                               #  4: transverse mass of Z boson (lepton system, mT_lep1+lep2)
                               #  5: geometric avarage of mT of photon and mT of Z boson
                               #  6: quadratic sum of Z mass and mT of the photon (mu^2=m_Z^2+mT_photon^2)
                               #  7: quadratic sum of dilepton mass and mT of photon (mu^2=m_lep1+lep2^2+mT_photon^2)
factor_central_scale = 1       #  relative factor for central scale (important for dynamic scales)
scale_variation  = 1           #  switch for muR/muF uncertainties (0) off; (1) 7-point (default); (2) 9-point variation
variation_factor = 2           #  symmetric variation factor; usually a factor of 2 up and down (default)

# scale settings for resummation
scale_res         =  91.1876   #  resummation scale
dynamic_scale_res =  1         #  dynamic resummation scale
                               #  0: fixed scale above
                               #  1: invariant mass (Q) of system (of the colourless final states)
factor_scale_res = 0.5         #  relative factor for central resummation scale (important for dynamic scale)
scale_variation_res  = 1       #  switch for resummation-scale uncertainties (0) off; (1) on (default);
variation_factor_res = 2       #  symmetric variation factor; usually a factor of 2 up and down (default)
\end{lstlisting}
}

\vspace{0.2cm}\noindent
\texttt{dynamic\_scale}\quad This parameter is present already in a fixed-order run to choose between the 
specified fixed renormalisation and factorisation scales (\texttt{scale\_ren}/\texttt{scale\_fact}) and dynamic ones, and we refer the reader to the information given section~5.1.1.2 of ref.~\cite{Grazzini:2017mhc} 
for details. In a resummed calculation it is important to only use sensible choices for dynamical 
scales, i.e.\ any choice that depends on QCD radiation must smoothly approach
the scale at Born level in the limit where the QCD radiation becomes soft or collinear.
In particular, this allows for any dynamical scale choice that only involves the Born level momenta
of the color singlets.

\vspace{0.2cm}\noindent
\texttt{dynamic\_scale\_res}\quad This parameter allows the user to choose between the specified fixed resummation scale (\texttt{scale\_res}) and a dynamic one. The dynamic scale, however,
is limited to the invariant mass of the colour singlet, which is also the recommended setting.

\vspace{0.2cm}\noindent
\texttt{factor\_central\_scale\_res}\quad A relative factor that multiplies the central resummation scale; particularly useful for a dynamic resummation scale. It is recommended to set it to a fraction of the invariant mass of the color singlet (setting \texttt{dynamic\_scale\_res = 1}) such as \texttt{0.25} or \texttt{0.5}.

\vspace{0.2cm}\noindent
\texttt{scale\_variation\_res}\quad Switch to turn on and off scale variations of the resummation scale.
If set to \texttt{1}, a variation up and down will be done while keeping the renormalisation and factorisation
at their central values. The total uncertainty is calculated as the envelope of the 
renormalisation and factorisation scale variations (1, 7, or 9 variations) and of the additional two variations of the resummation scale, for a total of 3, 9, or 11 variations.

\vspace{0.2cm}\noindent
\texttt{variation\_factor\_res}\quad This (integer) value determines by which factor 
with respect to the central resummation scale the scale variation is performed.

\newpage
\paragraph{Order-dependent run settings} 
\lstset{basicstyle=\tiny, frame=single}
{\tt
\begin{lstlisting}
# LO
run_LO          =  1           #  switch for LO cross section (1) on; (0) off
LHAPDF_LO       =  NNPDF30_lo_as_0118 #  LO LHAPDF set
PDFsubset_LO    =  0           #  member of LO PDF set
precision_LO    =  1.e-2       #  precision of LO cross section

# NLO(+NLL)
run_NLO         =  0           #  switch for NLO cross section (1) on; (0) off
add_RadISH_NLL  =  0           #  switch to add NLL RadISH resummation to NLO (1) on; (0) off
LHAPDF_NLO      =  NNPDF30_nlo_as_0118 #  NLO LHAPDF set
PDFsubset_NLO   =  0           #  member of NLO PDF set
precision_NLO   =  1.e-2       #  precision of NLO cross section
NLO_subtraction_method = 1     #  switch to use (2) qT subtraction (1) Catani-Seymour at NLO

# NNLO(+NNLL)
run_NNLO        =  0           #  switch for NNLO cross section (1) on; (0) off
add_RadISH_NNLL =  0           #  switch to add NNLL RadISH resummation to NNLO (1) on; (0) off
add_RadISH_N3LL =  0           #  switch to add N3LL RadISH resummation to NNLO (1) on; (0) off
LHAPDF_NNLO     =  NNPDF30_nnlo_as_0118 #  NNLO LHAPDF set
PDFsubset_NNLO  =  0           #  member of NNLO PDF set
precision_NNLO  =  1.e-2       #  precision of NNLO cross section
loop_induced    =  1           #  switch to turn on (1) and off (0) loop-induced gg channel

switch_qT_accuracy = 0         #  switch to improve qT-subtraction accuracy (slower numerical convergence) 
\end{lstlisting}
}

\vspace{0.2cm}\noindent
\texttt{add\_RadISH\_NLL}\quad Switch to turn on and off the NLL resummation through \textsc{RadISH} in the NLO run and the matching to NLO QCD. Available
for all \textsc{RadISH} observables (settings of \texttt{RadISH\_observable} below).

\vspace{0.2cm}\noindent
\texttt{add\_RadISH\_NNLL}\quad Switch to turn on and off the NNLL resummation through \textsc{RadISH} in the NNLO run and the matching to NNLO QCD. Available
for all \textsc{RadISH} observables (settings of \texttt{RadISH\_observable} below). Can 
only be turned on if \texttt{add\_RadISH\_N3LL} below is turned off.

\vspace{0.2cm}\noindent
\texttt{add\_RadISH\_N3LL}\quad Switch to turn on and off the N$^3$LL resummation through \textsc{RadISH} in the NNLO run and the matching to NNLO QCD. Available
only for transverse-momentum resummation of the color singlet (\texttt{RadISH\_observable = 1} below). Can only be turned on if \texttt{add\_RadISH\_NNLL} above is turned off.

\vspace{0.2cm}\noindent
\texttt{loop\_induced}\quad This switch is present already in a fixed-order run for certain processes 
(such as $ZZ$, $W^+W^-$, \ldots{}) where a loop-induced $gg$ contribution enters at the NNLO. In a 
resummation (for now) the switch has to be set to \texttt{0}, since the effectively LO-accurate 
$gg$ contribution would trivially enter the resummed observables under consideration.

\newpage
\paragraph{RadISH-specific settings.} 
\lstset{basicstyle=\tiny, frame=single}
{\tt
\begin{lstlisting}
RadISH_observable = 1          #  observable to be resummed by RadISH at the accuracy specified above
                               #  1: transverse-momentum of system (of the colourless final states)
                               #  2: transverse-momentum of the leading jet (no N3LL supported!);
                               #      jets MUST be defined like this:
                               #     "define_eta jet = 1.e99"
                               #     "define_y jet = 1.e99"
                               #     "n_observed_min jet = 0"
                               #     "n_observed_max jet = 99"
                               #     AND:
                               #     "jet_R_definition = 0"
                               #  3: phi-star angle in Drell-Yan process (ONLY available for ppeex02 and ppnenex02)
                               #  4: double-differential resummation of transverse-momentum of system & leading jet;
                               #     computes transverse-momentum of system with a jet-veto cut;
                               #     the jet veto MUST be defined like this:
                               #     "define_pT jet = value"
                               #     "define_eta jet = 1.e99"
                               #     "define_y jet = 1.e99"
                               #     "n_observed_min jet = 0"
                               #     "n_observed_max jet = 0"
                               #     AND:
                               #     "jet_R_definition = 0"
                               #     "jet_R = 0.4" or "jet_R = 0.8"
                               #  5: double-differential resummation of transverse-momentum of system & leading jet;
                               #     computes transverse-momentum of leading jet with a veto cut on the system;
                               #     the veto cut MUST be defined by setting RadISH_pTsystem_veto and jets like this:
                               #     "define_eta jet = 1.e99"
                               #     "define_y jet = 1.e99"
                               #     "n_observed_min jet = 0"
                               #     "n_observed_max jet = 99"
                               #     AND:
                               #     "jet_R_definition = 0"
                               #     "jet_R = 0.4" or "jet_R = 0.8"
modlog_p = 4                   #  order modified logarithms (integer from 1 to 5; 4 for pT/phi-star, and 5 for pTjet)
matching_scheme = 1            #  matching scheme used to combine the fixed order and resummation
                               #  0: additive
                               #  1: multiplicative (default)
number_of_events = 2000        #  number of radish events used for each Born phase space point
min_born_events_radish = 2000  #  minimal number of Born events used in each parallel run
RadISH_pTsystem_veto = 1.e99   #  veto on transverse momentum of system (used only for RadISH_observable = 5)
\end{lstlisting}
}

\vspace{0.2cm}\noindent
\texttt{RadISH\_observable}\quad This switch determines the observable resummed by \textsc{RadISH}.
  Currently the following options are supported:
  \begin{enumerate}
  \item transverse momentum of the colour singlet; 
  \item transverse momentum of the leading jet; requires to set the jet definition in \texttt{parameter.dat} with the following mandatory settings:
  \lstset{basicstyle=\scriptsize, frame=single}
{\tt
\begin{lstlisting}
# Jet algorithm
...
jet_R_definition = 0           #  (0) pseudo-rapidity (1) rapidity

...

# Jet cuts
...
define_eta jet = 1.e99         #  requirement on jet pseudo-rapidity (upper cut)
define_y jet = 1.e99           #  requirement on jet rapidity (upper cut)
n_observed_min jet = 0         #  minimal number of observed jets (with cuts above)
n_observed_max jet = 99        #  maximal number of observed jets (with cuts above)
\end{lstlisting}
}
  \item $\phs$ angle in Drell--Yan production~\cite{Banfi:2010cf}; only available when running $pp\to l^+l^-$ (\texttt{ppeex02}) or $pp\to \nu\bar\nu$ (\texttt{ppnenex02});
  \item transverse momentum of the colour singlet with a jet veto (double-differential resummation); 
  requires to set a jet veto in the settings for fiducial cuts in \texttt{parameter.dat} with the following mandatory settings:
  \lstset{basicstyle=\scriptsize, frame=single}
{\tt
\begin{lstlisting}
# Jet algorithm
...
jet_R_definition = 0           #  (0) pseudo-rapidity (1) rapidity
jet_R = ${jet-radius-value}    #  DeltaR

...

# Jet cuts
define_pT jet = ${veto-value}  #  requirement on jet transverse momentum (lower cut)
define_eta jet = 1.e99         #  requirement on jet pseudo-rapidity (upper cut)
define_y jet = 1.e99           #  requirement on jet rapidity (upper cut)
n_observed_min jet = 0         #  minimal number of observed jets (with cuts above)
n_observed_max jet = 0         #  maximal number of observed jets (with cuts above)
\end{lstlisting}
}
where only the value of the jet veto (\texttt{\$\{veto-value\}}) may be chosen freely between \texttt{1} and \texttt{1.e99}, a typical 
setting would be for instance\ \texttt{define\_pT jet = 30.}, and the value of the jet radius (\texttt{\$\{jet-radius-value\}}) may be chose either to be \texttt{0.4} or \texttt{0.8};\footnote{Please contact the authors if other values for the jet radius are desired.}
  \item transverse momentum of the leading jet with a veto on the transverse momentum of the color singlet (double-differential resummation); requires to set a veto on the transverse momentum of the color singlet via \texttt{RadISH\_pTsystem\_veto} between \texttt{1} and \texttt{1.e99}, a typical 
setting would be for instance\ \texttt{RadISH\_pTsystem\_veto = 30.}\,; and to set the jet definition in \texttt{parameter.dat} with the following mandatory settings:
  \lstset{basicstyle=\scriptsize, frame=single}
{\tt
\begin{lstlisting}
# Jet algorithm
...
jet_R_definition = 0           #  (0) pseudo-rapidity (1) rapidity
jet_R = ${jet-radius-value}    #  DeltaR

...

# Jet cuts
...
define_eta jet = 1.e99         #  requirement on jet pseudo-rapidity (upper cut)
define_y jet = 1.e99           #  requirement on jet rapidity (upper cut)
n_observed_min jet = 0         #  minimal number of observed jets (with cuts above)
n_observed_max jet = 99        #  maximal number of observed jets (with cuts above)
\end{lstlisting}
}
  \end{enumerate}

\vspace{0.2cm}\noindent
\texttt{modlog\_p}\quad This (integer) value sets the exponent $p$ in the modified logarithms defined in eq.~\eqref{eq:modlog}. Has to be chosen between \texttt{1} and \texttt{5}. We recommend a setting 
of \texttt{modlog\_p = 4} for the transverse momentum of the color singlet and $\phs$, and a setting of 
\texttt{modlog\_p = 5} for the transverse momentum of the leading jet.
  
\vspace{0.2cm}\noindent
\texttt{matching\_scheme}\quad Switch to choose between the additive matching scheme in eq.~\eqref{eq:additive} or the multiplicative matching in eq.~\eqref{eq:multiplicative1}. 
We recommend using the multiplicative matching for the various reasons discussed in section~\ref{sec:matching}.
In order to compare results obtained with both schemes for the same run, first complete a run
with one matching scheme; then choose the other matching scheme by modifying the file \texttt{parameter.dat} of the respective run inside the folder \texttt{input} and restart the combination (and matching) of the results with the command
  \begin{Shaded}
\begin{Highlighting}[]
  \ExtensionTok{\$  ./bin/run_process}\NormalTok{ run_name_resummation}\NormalTok{ --run_mode run_matching}
\end{Highlighting}
\end{Shaded}
\noindent If enabled (default) in the file \texttt{parameter.dat} the previous results will be saved in
a subfolder \texttt{saved\_result\_XX} inside the \texttt{result} folder of the respective run, 
where \texttt{XX} is an increasing number starting at \texttt{01}.

\vspace{0.2cm}\noindent
\texttt{number\_of\_events}\quad This (integer) value sets the number of events used for each 
Born event in the \textsc{RadISH} Monte Carlo implementation. 
We recommend to use a value between \texttt{2000} and \texttt{5000}, and not lower than \texttt{1000} to avoid numerical artefacts. A further increase of this number may be useful for dedicated runs with 
very fine binning at small transverse momenta, but one must bear in mind that this will slow 
down the computation.

\vspace{0.2cm}\noindent
\texttt{min\_born\_events\_radish}\quad This (integer) value sets the minimum number of Born events for each parallel run of the resummation with \textsc{RadISH}. 
We recommend to use a value between \texttt{2000} and \texttt{5000}, and not lower than \texttt{1000} to avoid numerical artefacts. 
In heavily parallelised runs, this ensures that each parallel run of the resummation with \textsc{RadISH} has a minimum number of Born events, so that the numerical integration of each of them converges sufficiently 
before combining all parallel runs.

\subsubsection{Settings in {\tt distribution.dat}}\label{distribution-settings}

In the file \texttt{distribution.dat} the user defines histograms for distributions which are filled during the run. Each distribution is represented by one block of certain parameters. We refer the reader to section~5.1.3 of ref.~\cite{Grazzini:2017mhc} for details.

In order to perform a resummation run
it is mandatory that there is at least one block with \texttt{distributiontype = RadISH\_observable}. 
This special \texttt{distributiontype} automatically selects the correct observable according 
to the setting of the \texttt{RadISH\_observable} inside the file \texttt{parameter.dat} of the respective run, 
described in the previous section. In the first version of \textsc{Matrix+RadISH} all 
other distributions, i.e.\ those with \texttt{distributiontype} different from \texttt{RadISH\_observable}, 
are ignored for the calculation of the resummation and the matching. Their fixed-order results 
will be calculated and provided in the \texttt{result} folder as in an ordinary fixed-order run though. Thus,
the resummation and matching are evaluated only for distribution blocks with \texttt{distributiontype = RadISH\_observable}, which corresponds to the observable selected for the resummation.

The starting point, selected either via \texttt{startpoint} or via the first entry of \texttt{edges}, 
of any distribution block using \texttt{distributiontype = RadISH\_observable} must be set to \texttt{0} in order 
to warrant the correct matching of fixed-order prediction and resummation.
The user is free to add an arbitrary number of distribution blocks with \texttt{distributiontype = RadISH\_observable}, provided that the \texttt{distributionname} of each distribution block is different.
Below, we provide an example for the case of a regular and an irregular binning:

\newpage
\lstset{basicstyle=\small, frame=single}
{\tt
\begin{lstlisting}
distributionname  =  RadISH_observable_regular
distributiontype  =  RadISH_observable
startpoint        =  0.
endpoint          =  1000.
binnumber         =  200

distributionname  =  RadISH_observable_irregular
distributiontype  =  RadISH_observable
binningtype       =  irregular
edges             =  0.:2.:5.:10.:50.:100.:200.:500.:800.:1000.:50000.
\end{lstlisting}
}
\noindent We stress again that any additional fixed-order distribution alongside these histograms may be added as in a usual \textsc{Matrix} run, but that they will not be filled for the resummation 
and the matching of the calculation.


\bibliography{biblio}

\end{document}